\pdfoutput=1
\documentclass[12pt,a4paper]{amsart}
\usepackage[a4paper,inner=2.5cm,outer=2.5cm,top=2.5cm,bottom=2.5cm]{geometry}

\usepackage{amsmath,amssymb,amsthm,enumerate,mathtools,stmaryrd}
\usepackage{hyperref}
\hypersetup{colorlinks=true,linkcolor=blue,citecolor=teal,filecolor=magenta,urlcolor=cyan}
\usepackage{longtable}

\usepackage{makecell}
\usepackage{graphicx}

\usepackage{float}
\usepackage[matrix,arrow]{xy}

\usepackage{extarrows}

\usepackage{xcolor}

\def\xvital{\textsf{key}}
\def\yvital{$\vee$-\textsf{key}}
\def\regular{{non-special}}
\def\nonregular{{special}}

\usepackage{url}

\usepackage[backend=bibtex,style=alphabetic,sorting=nyt,isbn=false,url=false,doi=true,maxalphanames=10,minalphanames=4,mincitenames=4,maxcitenames=10,minnames=4,maxnames=10,giveninits=true,maxbibnames=99]{biblatex}
\theoremstyle{plain}
\newtheorem{theorem}{Theorem}[section]
\newtheorem{proposition}[theorem]{Proposition}
\newtheorem{corollary}[theorem]{Corollary}
\newtheorem{lemma}[theorem]{Lemma}

\theoremstyle{definition}
\newtheorem{definition}[theorem]{Definition}

\makeatletter
\@addtoreset{proofpart}{theorem}
\makeatother
\theoremstyle{remark}
\newtheorem{remark}[theorem]{Remark}

\def\xvital{\textsf{key}}
\def\yvital{$\vee$-\textsf{key}}
\def\regular{{non-special}}
\def\nonregular{{special}}

\newcommand{\Z}{\mathbb{Z}}
\newcommand{\C}{\mathbb{C}}
\newcommand{\bW}{\mathbb{W}}

\newcommand{\cS}{\mathcal{S}}
\newcommand{\cA}{\mathcal{A}}

\newcommand{\cP}{\mathcal{P}}

\newcommand{\cW}{\mathcal{W}}
\newcommand{\cT}{\mathcal{T}}
\DeclareMathOperator{\Aut}{Aut}

\newcommand{\res}{\mathop{\rm res}}

\newcommand{\restr}[2]{\mathop{\big\lfloor_{{#1}\to {#2}}}}
\newcommand{\set}[1]{\llbracket {#1} \rrbracket}

\newcommand{\ii}{\mathrm{i}}
\newcommand{\np}{\mathsf{np}}
\newcommand{\Kr}{\mathsf{Kr}}
\newcommand{\fS}{\mathfrak{S}}

\newcommand{\fA}{\mathfrak{A}}
\newcommand{\fB}{\mathfrak{B}}
\renewcommand{\Im}{\mathop{\mathrm{Im}}}

\newcommand{\cg}{\mathfrak{g}}

\title[KP integrability of non-perturbative differentials]
{KP integrability of non-perturbative differentials}

\author[A.~Alexandrov]{A.~Alexandrov}
\address{A.~A.: Center for Geometry and Physics, Institute for Basic Science (IBS), Pohang 37673, Korea
}
\email{alex@ibs.re.kr}

\author[B.~Bychkov]{B.~Bychkov}
\address{B.~B.: Department of Mathematics, University of Haifa, Mount Carmel, 3498838, Haifa, Israel}
\email{bbychkov@hse.ru}

\author[P.~Dunin-Barkowski]{P.~Dunin-Barkowski}
\address{P.~D.-B.: Faculty of Mathematics, National Research University Higher School of Economics, Usacheva 6, 119048 Moscow, Russia; HSE--Skoltech International Laboratory of Representation Theory and Mathematical Physics, Skoltech, Nobelya 1, 143026, Moscow, Russia; and ITEP, 117218 Moscow, Russia}
\email{ptdunin@hse.ru}

\author[M.~Kazarian]{M.~Kazarian}
\address{M.~K.: Faculty of Mathematics, National Research University Higher School of Economics, Usacheva 6, 119048 Moscow, Russia; and Center for Advanced Studies, Skoltech, Nobelya 1, 143026, Moscow, Russia}
\email{kazarian@mccme.ru}

\author[S.~Shadrin]{S.~Shadrin}
\address{S.~S.: Korteweg-de Vries Institute for Mathematics, University of Amsterdam, Postbus 94248, 1090GE Amsterdam, The Netherlands}
\email{S.Shadrin@uva.nl}

\begin{document}

\begin{abstract}
	We prove the KP integrability of non-perturbative topological recursion, which can be considered as a formal $\hbar$-deformation  of the Krichever construction of algebro-geometric solutions of the KP hierarchy. This property goes back to a 2011 conjecture of Borot and Eynard.
\end{abstract}

\maketitle

\tableofcontents

\section{Introduction}

The celebrated Novikov conjecture on the Riemann--Schottky problem stated that the Jacobians of smooth algebraic curves are exactly those indecomposable principally polarized abelian varieties  whose $\Theta$ functions are KP integrable. It prompted an enormous development on the edge of algebraic geometry and integrable systems that has appeared to be very fruitful for both areas, and it was resolved by Krichever and Shiota (see~\cite{Krichever-main},~\cite{Shiota}, and also~\cite{KricheverShiota} for their joint survey of this topic). A far-reaching outcome for the theory of integrable systems is a huge supply of the so-called algebro-geometric solutions of the KP hierarchy constructed by Krichever~\cite{Krichever-main} that can be associated to a choice of a compact Riemann surface $\Sigma$,
a point $p\in\Sigma$ with a local coordinate at it, and a few more parameters. Remarkably, the Krichever construction admits a system of deformations (preserving the KP integrability property) that depend on an additional choice of two meromorphic functions (or differentials) on $\Sigma$. Existence of these deformations was conjectured by Borot and Eynard in~\cite{BorEyn-AllOrderConjecture}, and the purpose of this paper is to give a proof of the Borot--Eynard conjecture.

While the implications of the existence of these deformations of the Krichever algebro-geometric solutions of the KP hierarchy in the context of the Riemann--Schottky-type problems are yet to be studied, it is worth to mention that the ideas that led Borot and Eynard to their conjecture are closely connected to a completely different corner of mathematics, and subsequently they got applications to knot theory~\cite{BorEyn-knots}. In a nutshell, the KP integrability in this latter case is a step towards understanding the strong ties that connect the hyperbolic geometry of the complement of a knot and its quantum invariants coming from the Chern--Simons theory (the full correspondence is the subject of the so-called AJ conjecture).

\subsection{Notation} Throughout the paper we use the following notation:

\begin{itemize}
	\item
Let $\set n$ be the set of indices $\{1,\dots,n\}$. For each $I\subset \set n$ let $z_I$ denote the set of formal variables / points on a Riemann surface / functions (respectively, depending on the context) indexed by $i\in I$.

\item

For the $n$-th symmetric group $\fS_n$ its subset of cycles of length $n$ is denoted by $C_n$. Let $\det^\circ (A_{ij})$ denote the connected determinant of a matrix $(A_{ij})$ given in the case of an $n\times n$ matrix by
\begin{align}
	{\det}^\circ (A_{ij})  \coloneqq (-1)^{n-1}  \sum_{\sigma \in C_n} \prod_{i=1}^n A_{i,\sigma(i)}.
\end{align}

\item
The operator $\restr{z}{z'}$ is the operator of restriction of an argument $z$ to the value $z'$, that is, $\restr{z}{z'}f(z) = f(z')$. When we write $\restr{}{z}$ we mean that the argument of a function or a differential to which this operator applies is set to $z$, without specifying the notation for the former argument.

\item
In order to shorten the notation we often omit the summation over an index that runs from $1$ to $\cg$. For instance, $\theta_{*}\eta \coloneqq \sum_{i=1}^\cg \theta_{*,i}\eta_i$, $\int_{\fB} \partial_w \coloneqq \sum_{i=1}^\cg \int_{\fB_i} \partial_{w_i}$, $\eta(z) \partial_w \coloneqq \sum_{i=1}^\cg \eta_i(z) \partial_{w_i}$, $k^T\cT k \coloneqq \sum_{i,j=1}^\cg k_i \cT_{ij} k_j$, $b^Tw \coloneqq \sum_{i=1}^\cg b_iw_i$, etc.

\item
Let $\cS(z)$ denote $z^{-1}(e^{\frac z2}-e^{-\frac z2} )$.

\end{itemize}

\subsection{KP integrability for a system of differentials}\label{sec:KP}
We study KP integrability as a property of a system of differentials, see e.g.~\cite{Kawamoto,Ooguri}, or, more recently,~\cite{ABDKS3}.

Consider a connected smooth Riemann surface $\Sigma$. Let $\omega_n$, $n\geq 1$, be a system of symmetric meromorphic $n$-differentials such that $\omega_n$ is regular on all diagonals for $n>2$, and $\omega_2$ has a pole of order two on the diagonal with biresidue~$1$. Define the bi-halfdifferential
\begin{align}\label{eq:Omega1}
&
\Omega_1(z^+,z^-)=
\\ \notag &
\frac{\sqrt{d\chi(z^+)}\sqrt{d\chi(z^-)}}{\chi(z^+)-\chi(z^-)}
\;\exp\left(\sum_{m=1}^\infty\frac1{m!}\Bigl(\int_{z^-}^{z^+}\Bigr)^m
\left(\omega_m(z_{\set{m}})-\delta_{m,2}\frac{d\chi(z_1)d\chi(z_2)}{(\chi(z_1)-\chi(z_2))^2}\right)\right),
\end{align}
where $\chi$ is any meromorphic function, which can serve as a local coordinate in the vicinity of the points $z^+$ and $z^-$; the result is actually independent of its choice.

\begin{definition} \label{def:KP-1} We say that $\{\omega_n\}_{n\geq 1}$ is KP integrable, if the following determinantal identity holds:
\begin{equation}
\omega_n(z_{\set n})=\det\nolimits^\circ\|\Omega_1(z_i,z_j)\|,\quad n\ge2.
\end{equation}
\end{definition}

The diagonal entries of the matrix $\|\Omega_1(z_i,z_j)\|$ are not defined but the connected determinant does not involve them for $n\ge2$.

\begin{remark}
The differential $\omega_1$ can also be recovered from $\Omega_1$ by
\begin{equation}
\omega_1(z_1)=\restr{z_2}{z_1}\Bigl(\Omega_1(z_1,z_2)-\frac{\sqrt{d\chi(z_1)}\sqrt{d\chi(z_2)}}{\chi(z_1)-\chi(z_2)}\Bigr),
\end{equation}
for any choice of a local coordinate~$\chi$, but this equality holds independently of KP integrability.

In fact, a choice of $\omega_1$ does not affect KP integrability and we could put simply $\omega_1=0$ and to drop the summand with $m=1$ in the formula for $\Omega_1$. With this modification of $\Omega_1$ the determinantal formula still holds. Note, however, that it does influence the relation between the actually given $\omega_1$ and $\Omega_1$, so we stick to the form of $\Omega_1$ as above.
\end{remark}

	A point $o\in\Sigma$ is called \emph{regular} for the system of differentials $\{\omega_n\}$ if $\omega_n-\delta_{n,2}\frac{d z_1d z_2}{(z_1-z_2)^2}$ is regular at $(o,\dots,o)\in\Sigma^n$ for all $n\in\Z_{>0}$, where~$z$ is a local coordinate on $\Sigma$ near~$o$.
To any regular point $o\in\Sigma$ and a local coordinate $z$ at this point we can associate a formal infinite power series $\tau=\tau_{o,z}(t_1,t_2,\dots)$ defined by the power expansion in the local coordinate~$z$ of the following equality
\begin{equation}\label{eq:n-differential}
\omega_n(z_{\set n})=\restr{t}{0}\delta_{z_1}\dots\delta_{z_n}\log(\tau)+\delta_{n,2}\frac{dz_1dz_2}{(z_1-z_2)^2},
\qquad \delta_z=\sum_{k=1}^\infty z^{k-1}dz\; \frac{\partial}{\partial t_k}.
\end{equation}
Then the KP integrability for a system of differentials~$\{\omega_n\}$ is equivalent to the condition that $\tau$ is a KP tau function for at least one choice of regular~$o$ and~$z$, and then it is automatically a KP tau function for any choice of regular~$o$ and~$z$~\cite[Theorem 2.3]{ABDKS3}.

\medskip
While discussing KP integrability for a system of differentials $\{\omega_n\}$, it is useful to consider an extended set of $n|n$ differentials (in fact, half-differentials) $\Omega_n$ and $\Omega_n^\bullet$ defined as follows. Let $B$ be any symmetric bidifferential with a second order pole on the diagonal with biresidue one,
 in particular, one can set $B=\omega_2$ or simply $B=\frac{dz_1dz_2}{(z_1-z_2)^2}$ for some local coordinate~$z$. In the global situation, when~$\Sigma$ is compact, one usually takes for $B$ the Bergman kernel which is normalized by vanishing $\mathfrak A$-periods.  Consider also the form~$E$ defined by
\begin{align}
	\frac{1}{E(z^+,z^-)}=\frac{\sqrt{d\chi(z^+)}\sqrt{d\chi(z^-)}} {\chi(z^+)-\chi(z^-)}
		\exp\Bigg(\frac 12 \int_{z^-}^{z^+}\int_{z^-}^{z^+}\Bigl( B(z_1,z_2) - \frac{d\chi(z_1)d\chi(z_2)}{(\chi(z_1)-\chi(z_2))^2}\Bigr)
		\Bigg) .
\end{align}
In particular, the right hand side is independent of the choice of a meromorphic function $\chi$ that can be considered as a local coordinate in the vicinity of $z^+$ and $z^-$ (see~\cite[Equation (3.10)]{ABDKS3}), but it depends on the choice of $B$, and for a specific $B$ coincides with the prime form, see Section~\ref{S2.2}. With this notation, the so called disconnected extended $n|n$ differentials are defined by
\begin{align} \label{eq:Omega-bullet-expint}
	\Omega^\bullet_n(z^+_{\set n},z^-_{\set n}) = 	&  \prod_{1\leq k<l\leq n} \frac{E(z^+_k,z^+_l)E(z^-_k,  z^-_l)}{E(z^+_k, z^-_l)E(z^-_k,z^+_l)}
\prod_{i=1}^n \frac{1%
}{E(z^+_i, z^-_i)}
	\\ \notag
	&
\times\exp\Bigg(
			\sum_{\scriptstyle\substack{m_1,\dots,m_n\geq 0 \\ \sum_{i=1}^n m_i=m\geq 1}}
			\prod_{i=1}^n \frac{1}{m_i!} \Big(\int_{z_i^-}^{z_i^+} \Big)^{m_i} (\omega_m-\delta_{m,2}B)
		\Bigg).
\end{align}
Their connected counterparts are given by the following formula:
\begin{align}\label{eq:Omega-expint}
	\Omega_n(z^+_{\set{n}},z^-_{\set n})  \coloneqq \sum_{\ell=1}^n \frac{(-1)^{\ell-1}}{\ell} \sum_{\scriptstyle\substack{I_1\sqcup \cdots \sqcup I_\ell = \set n \\ \forall j I_j\not=\emptyset}} \prod_{j=1}^\ell \Omega_{|I_j|}^{\bullet}(z^+_{I_j},z^-_{I_j}).
\end{align}
One can check that the definition of these symmetric bi-halfdifferentials does not depend on a choice of~$B$. In particular, expressions~\eqref{eq:Omega-bullet-expint}--\eqref{eq:Omega-expint} for $\Omega_1=\Omega^\bullet_1$ are equivalent to~\eqref{eq:Omega1}. Notice that the poles on the diagonals $z_i^+=z_i^-$ cancel out for the connected $n|n$ differentials for $n\ne1$, and we have
\begin{equation}
\omega_n (z_{\set n}) = \restr{z^+_{\set n} }{z_{\set n}}\restr{z^-_{\set n}}{z_{\set n}}\Omega_n(z^+_{\set n}, z^-_{\set n}),\quad n\ge2.
\end{equation}

The following statement is a well known reformulation of KP integrability (cf.~\cite[Theorem 4.4]{ABDKS3}, where it is stated in almost the same notation, or,  %
e.g., \cite{eynard2024hirotafaygeometry}):

\begin{proposition} \label{prop:equiv-KP} A system of differentials $\{\omega_n\}_{n\geq 1}$ is KP integrable if and only if the following determinantal identities for extended differentials hold true:
\begin{align}
\Omega^\bullet_n(z^+_{\set n},z^-_{\set n})&=\det\|\Omega_1(z^+_i,z^-_j)\|,\\
\Omega_n(z^+_{\set n},z^-_{\set n})&=\det\nolimits^\circ\|\Omega_1(z^+_i,z^-_j)\|.
\end{align}
\end{proposition}

Note that Definition~\ref{def:KP-1} can be then recovered as a special case of the second equation by taking $z_i^\pm\to z_i$, $i=1,\dots,n$.   
The following remarks regarding the definitions of $\Omega^\bullet_n$ and $\Omega_n$ are in order:

\begin{remark} Equations~\eqref{eq:Omega1} and \eqref{eq:Omega-bullet-expint} contain an infinite summation under the exponent. In order to avoid problems with convergence, we either have to interpret this formula as expansion in some suitable local coordinate, or assume some extra expansion of $\omega_n$ in a formal parameter $\hbar$ such that the coefficients of expansion in $\hbar$ are finite expressions.
\end{remark}

\begin{remark} Working with half-differentials we have to choose some square roots of the canonical line bundle on $\Sigma$ (this concerns both $E$ and $\Omega_n$). In the constructions below in terms of the $\Theta$ functions this choice is dictated by the choice of odd theta characteristics.
\end{remark}

An example of a KP integrable system of differentials is given by the so-called Krichever differentials~\cite{Krichever-main}, which we briefly recall below. Another example of a KP integrable system of differentials is any system of differentials produced by topological recursion on a genus $0$ spectral curve (we recall the definition below), see~\cite{alexandrov2024topologicalrecursionrationalspectral,alexandrov2024degenerateirregulartopologicalrecursion}.

Finally, our main result in this paper is that the former two constructions can be merged into a system of the so-called non-perturbative differentials of topological recursion, proposed and further studied in~\cite{EynardMarino,BorEyn-AllOrderConjecture,BorEyn-knots,eynard2024hirotafaygeometry}, which we also prove to be KP integrable, as it was conjectured by Borot and Eynard in~\cite{BorEyn-AllOrderConjecture}.

\medskip
We conclude this section with the following remark (not concerning KP integrability directly). For a given system of differentials $\{\omega_n\}$, along with the differentials $\Omega^\bullet_n$ and $\Omega_n$ we will use also their certain specializations. Namely, given additionally two functions~$x$ and~$y$, we can rewrite the connected $n|n$ half-differentials in different coordinates near the diagonal using the substitution $z_i^\pm = e^{\pm \frac{u\hbar}{2}\partial_{x_i}} z_i$, where $x_i\coloneqq x(z_i)$, $y_i\coloneqq y(z_i)$, simultaneously promoting them to the so-called extended $n$-differentials:
\begin{align}\label{eq:bW-general}
\bW_n(z_{\set n}, u_{\set n}) &\coloneqq  \Big(\prod_{i=1}^n e^{u_i\bigl(\cS(u_i\hbar\frac{d}{dx_i})-1\bigr) y_i}\Big)
\restr{z_i^\pm}{e^{\pm \frac{u_i\hbar}{2}\partial_{x_i}} z_i} \Omega_n(z^+_{\set{n}},z^-_{\set n}),
\\\label{eq:cW-general}	
\cW_n(z,u;z_{\set n}) & \coloneqq \bW_{n+1}(z,z_{\set n}, u, u_{\set n})\big\vert_{u_{\set n} = 0}.
\end{align}
Note that $\omega_n$ is an obvious specialization of those:
\begin{align}
	\omega_n (z_{\set n}) =\restr{u_{\set n}}{0} \left(\bW_n(z_{\set n}, u_{\set n})-\delta_{n,1}\frac{dx_1}{u_1\hbar}\right).
\end{align}

The point is that there are explicit combinatorial formulas expressing $\Omega_n$, $\bW_n$, and $\cW_n$ in terms of $\omega$-differentials with summation over certain graphs. We review and slightly revisit these formulas in %
Section~\ref{sec:constr-and-prop}.
Here we mention just a formula for $\cW_n$:
	\begin{align}\label{eq:cW1}
		\cW_n(z,u;z_{\set n})&=\frac{dx}{u\hbar}e^{\cT_0(z,u)}
	\sum_{\substack{\set{n}=\sqcup_{\alpha} J_\alpha
			\\J_\alpha\ne\emptyset}}
	\prod_{\alpha}\cT_{|J_\alpha|}(z,u;z_{J_\alpha}),
	\end{align}
	where
	\begin{align}\label{eq:cT1x} 
		& \cT_n(z,u;z_{\set n})\coloneqq 
		\\ \notag &
		\sum_{k=1}^\infty\frac1{k!}\prod_{i=1}^k
		\Bigl(\restr{\tilde z_{i}}{z}u\hbar\cS\big(u\hbar\frac{d}{d\tilde x_i}\big)\frac{1}{d\tilde x_{i}}\Bigr)
		\Bigl(\omega_{k+n}(\tilde z_{\set{k}},z_{\set{n}})-
		\delta_{n,0}\delta_{k,2}\frac{d\tilde x_{1}d\tilde x_{2}}{(\tilde x_{1}-\tilde x_{2})^2}\Bigr)
		\\\notag&+\delta_{n,0}u\bigl(\cS\big(u\hbar\frac{d}{dx}\big)-1\bigr)y.
	\end{align}

\subsection{Topological recursion} Topological recursion of Chekhov, Eynard, and Oran\-tin \cite{CEO,EO-1st} associates a system of meromorphic differentials $\omega^{(g)}_n$, $g\geq 0$, $n\geq 1$, $2g-2+n\geq 0$, to input data that consists of a Riemann surface $\Sigma$ and a finite set of points $\cP\subset \Sigma$, two meromorphic functions $x$ and $y$ on $\Sigma$  such that $\restr{}{q}dx = 0$ and $\restr {} {q} {dy}\not= 0$ for each $q\in \cP$ and a bi-differential $B$ with the double pole on the diagonal with biresidue $1$.
It has its origin in the computation of the cumulants of the matrix models, and by now it has multiple striking applications in algebraic geometry, enumerative combinatorics, and mathematical physics.

In the local setup~\cite{DOSS} one can assume that $\Sigma$ is just a union of disjoint discs around points in $\cP$. In the global setup one assumes that $\Sigma$ is a compact Riemann surface, and $B$ is the Bergman kernel. But even in the last case it is still sufficient to have $dx$ and $dy$ defined in some neighborhood of the points in $\cP$: the resulting differentials $\omega^{(g)}_n$ of topological recursion are globally defined meromorphic.

We follow~\cite{alexandrov2024degenerateirregulartopologicalrecursion}, though for the purposes of this paper we restrict ourselves to the case of a compact Riemann surface $\Sigma$ with a fixed choice of $\fA$ and $\fB$ cycles and the Bergman kernel $B$ normalized on $\fA$ cycles.
Let $dx,dy$ be two meromorphic differentials on $\Sigma$.
For each point $q\in\Sigma$ we consider the local expansions of $dx$ and $dy$ in some local coordinate $z$,
	\begin{align}
		dx=a\,z^{r-1}(1+O(z))dz,\quad dy=b\,z^{s-1}(1+O(z))dz,\quad a,b\ne0,\ r,s\in\Z,	
	\end{align}
and we say that the point $q$ is \regular{} if either $r=s=1$ or $r+s\le0$, and \nonregular{} otherwise.

\begin{definition}
The initial data of generalized topological recursion is a tuple $(\Sigma,dx,\allowbreak dy, \allowbreak \cP)$, where
$\cP\sqcup \cP^\vee$ is an arbitrarily chosen split of the set of \nonregular{} points.
\end{definition}

We call the points of $\cP$ (respectively, $\cP^\vee$) \xvital{}-points (respectively, \yvital{}-points).

\begin{definition} \label{def:TRgeneral}
	The differentials of generalized topological recursion $\omega^{(g)}_n$, $g\geq 0$, $n\geq 1$, $2g-2+n\geq 0$, for the initial data $(\Sigma,dx,dy,\cP)$ are defined by
	$\omega^{(0)}_2=B$
	and for $2g-2+n>0$ they are given by
	\begin{equation}
		\label{eq:newproj}
		\omega^{(g)}_{n}(z,z_{\set{n-1}})=\sum_{q\in\cP}\res\limits_{\tilde z=q} 
		\bigg(-\restr{z}{\tilde z}\sum_{r\ge1}\Bigl(-d\frac{1}{dy}\Bigr)^r[u^r]\cW^{(g)}_{n-1}(z,u;z_{\set {n-1}})
		\bigg)\int^{\tilde z}B(\cdot,z),
	\end{equation}
where the differentials $\cW_n=\sum_{g=0}^\infty\hbar^{2g-2+n}\cW^{(g)}_n$	are computed by~\eqref{eq:cW1} for the system of differentials $\omega_n=\sum_{\scriptstyle \substack{g\ge0\\(g,n)\ne(0,1)}} \hbar^{2g-2+n}\omega^{(g)}_{n}$, $x=x(z)$, $y=y(z)$.
\end{definition}

\begin{remark} This definition might look intimidating without the context that led to it; we refer the reader to~\cite{alexandrov2024degenerateirregulartopologicalrecursion} for a full discussion. In particular, it indeed recursively defines $\omega^{(g)}_n$, $2g-2+n>0$, which turn out to be symmetric meromorphic differentials with the poles only at the $\xvital{}$-points.
\end{remark}

\medskip
Yet another viewpoint to topological recursion is provided by the loop equations. They are applied if $dy$ is holomorphic and non-vanishing at each \xvital{}-point while $dx$ gets simple zeros at these points. %
We refer to this situation as \emph{the standard setup} of topological recursion.

\begin{definition} We say that a system of differentials $\{\omega^{(g)}_n\}$, $g\geq 0$, $n\geq 1$, $2g-2+n\geq 0$, satisfy the loop equations at the given point $q\in\cP$ if for any $k\ge0$ the following relation holds true:
	\begin{equation} \label{eq:(k-1)-loop}
		[u^k]e^{u y}\cW^{(g)}_n(z,u)\in\Xi_q,
	\end{equation}
	where the space $\Xi_q$ is spanned by meromorphic differentials of the kind $\bigl(d\frac{1}{dx}\bigr)^j\alpha$ where $j\ge 0$ and $\alpha$ is holomorphic.
\end{definition}

\begin{remark} In the standard setup it is sufficient to consider only $k=0$ and $k=1$ (the so-called linear and quadratic loop equations), see~\cite[Section 5]{ABDKS1}. However, we keep including all  $k\geq 0$ in the parts referring to the standard setup throughout the paper as it allows an immediate generalization of the relevant parts to the case when $dx$ possibly has higher order zeros at \xvital{}-points.
\end{remark}

\begin{lemma}[\cite{alexandrov2024degenerateirregulartopologicalrecursion}] \label{lemma:CEO-compatibility} In the standard setup
	the definition of the generalized topological recursion is equivalent to the following:
\begin{itemize}
	\item $\omega^{(0)}_2=B$;
	\item for $2g-2+n>0$ the differentials $\omega^{(g)}_n$ have poles only at the $\xvital{}$-points;
	\item $\{\omega^{(g)}_n\}$ satisfy the loop equations at the $\xvital{}$-points;
	\item they satisfy the projection property:
	\begin{equation}
		\label{eq:oldproj}
		\omega_{n}(z,z_{\set{n-1}})=\sum_{q\in\cP}\res\limits_{\tilde z=q}
		\omega_{n}(\tilde z,z_{\set{n-1}})\int^{\tilde z}B(\cdot,z).
	\end{equation}
\end{itemize}
In the global setup, that is, if $\Sigma$ is compact, there is a canonical choice of~$B$ dictated by requirement of vanishing its $\mathfrak A$ periods. For this choice of $B$, the projection property is also equivalent to the condition that the globally defined differentials $\omega^{(g)}_n$ have also vanishing $\mathfrak{A}$ periods.
\end{lemma}

\medskip
If we assume that $dx$ and $dy$ are globally defined on a compact Riemann surface $\Sigma$, one can trace the effect of the swap of $x$ and $y$ in the initial data of topological recursion on the resulting differentials. It is a powerful technique developed in a number of papers \cite{borot2023functional,hock2022xy,hock2022simple,ABDKS1}, and its natural development leads to a vast generalization and a revision of the definition of topological recursion, which was developed through a sequence of papers in~\cite{hock2023xy,ABDKS-logTR-xy,ABDKS-log-sympl} and got its final form in~\cite{alexandrov2024degenerateirregulartopologicalrecursion}.

\medskip

In this paper we use some techniques developed in~\cite{ABDKS3,alexandrov2024topologicalrecursionrationalspectral,alexandrov2024degenerateirregulartopologicalrecursion}, and we refer the reader to the corresponding parts of these papers when appropriate:
\begin{itemize}
	\item The $x-y$ swap action mentioned above and the fact that it preserves KP integrability of a system of differentials \cite{ABDKS3}.
	\item In the standard setup, there is a deformation formula that captures the effect of the infinitesimal change of $dy$ near the $\xvital{}$-points; it is proved that it preserves the KP integrability \cite{alexandrov2024topologicalrecursionrationalspectral}.
	\item Compatibility of the generalized topological recursion with particular type of limit behavior of the input data \cite{alexandrov2024degenerateirregulartopologicalrecursion}.
\end{itemize}

\subsection{Non-perturbative differentials}

The following result was proved in~\cite{ABDKS3,alexandrov2024topologicalrecursionrationalspectral} for the standard setup of topological recursion and extended in~\cite{alexandrov2024degenerateirregulartopologicalrecursion} for generalized topological recursion:

\begin{proposition} The system of differentials $\omega_n$, $n\geq 1$, constructed by (generalized) topological recursion is KP integrable if and only if the underlying Riemann surface $\Sigma$ has genus $0$.
\end{proposition}

For the Riemann surfaces of higher genus one has to deal with the so-called \emph{non-perturbative differentials} $\omega^\np_n$, $n\geq 1$, according to the conjecture of~\cite{BorEyn-AllOrderConjecture}. The non-perturbative differentials were initially introduced in~\cite{EynardMarino} in order to achieve the so-called background independence and modularity of the associated partition function. In~\cite{BorEyn-knots} they are related to the quantum invariants of knots.

The non-perturbative differentials can be expressed as formal power series in $\hbar$ of non-topological type, $\omega_n^{np} = \sum_{d=0}^\infty \hbar^d \omega_n^{np,\langle d \rangle}$, with the leading terms given by the so-called Krichever differentials that are the cornerstones of Krichever's construction of the finite-zone solutions of the KP equations.

\begin{remark}
Since we have this expansion in $\hbar$, the adjective ``non-perturbative'' might be a bit misleading, but it is by now the standard terminology in this area.	
\end{remark}

The non-perturbative differentials are the main objects of study in this paper. Their definition requires some preparation, so in Section~\ref{sec:JacobiansAndThetas} we briefly summarize the standard facts on $\Theta$ functions and recall the construction of the Krichever differentials, and in Section~\ref{sec:np-definitions} we merge the construction of the Krichever differentials with the input data of the generalized topological recursion discussed above in order to introduce the non-perturbative differentials and associated $n|n$ %
half-differentials.

The main results about the non-perturbative differentials which are the main results of the paper are presented in the rest of Section~\ref{sec:constr-and-prop} and can be summarized as follows:
\begin{itemize}
	\item In the standard setup, they can be described via loop equations (Theorem~\ref{thm:loop-equations}) and a certain deformation of the projection property (Theorem~\ref{thm:projectionproperty}), which makes them the subject of a suitable generalization of the so-called blobbed topological recursion~\cite{BS-blobbed}.
	\item In the generalized setup, the non-perturbative differentials obey the standard formulas of the $x-y$ swap (Theorem~\ref{thm:xyswap}). These formulas are known to preserve the KP integrability of a system of differentials.
	\item In the standard setup, the non-perturbative differentials obey the standard formulas of the $dy$-deformations (Theorem~\ref{thm:Deformation-y}). These formulas are also known to preserve the KP integrability of a system of differentials.
	\item The previous two results 
	allow to prove the KP integrability of the system of non-perturbative differentials (Theorem~\ref{thm:KP-integrable}). We also present a version of this statement with some extra free parameters (Corollary~\ref{cor:KP-deformed}) that can be tuned to achieve the background independence and re-introduce the KP times that are otherwise omitted in our approach to KP integrability.
\end{itemize}

\begin{remark} Theorem~\ref{thm:KP-integrable} resolves the conjecture of Borot and Eynard in~\cite{BorEyn-AllOrderConjecture}. We discuss in Section~\ref{sec:discussion-formulations} the precise statements in~\cite{BorEyn-AllOrderConjecture} and their interrelation in our notation.
\end{remark}

\subsection{Acknowledgments} We thank G. Borot and B. Eynard for useful discussions and encouragement to apply our methods to this problem. Substantial part of this project was completed during the stay of the authors at the CGP IBS in Pohang, which we thank for hospitality. We also thank the anonymous referees for valuable remarks.

A.~A. was supported by the Institute for Basic Science (IBS-R003-D1). B.~B. was supported by the ISF Grant 876/20. B.~B., P.~D.-B., and M.~K. were supported by the Russian Science Foundation (grant No. 24-11-00366). S.~S. was supported by the Dutch Research Council grant OCENW.M.21.233.

\subsection{A further development} \label{sec:Blobs-Gen} Since this paper was first submitted, we found a new proof of the Borot--Eynard conjecture based on completely different ideas, see~\cite{alexandrov2025blobbedtopologicalrecursionkp}. The principal difference is the deformation techniques used in the present paper (where the Krichever differentials are fixed leading terms and we move across a variety of KP integrable deformations of the Krichever differentials) and in the \emph{op.~cit.} (where the Krichever differentials themselves are treated as instances of the so-called blobs~\cite{BS-blobbed} and get deformed through their restrictions to some open disks).

\section{Necessary facts about Jacobians and \texorpdfstring{$\Theta$}{Theta} functions}
\label{sec:JacobiansAndThetas}

\subsection{\texorpdfstring{$\Theta$}{Theta} functions}

We fix a smooth algebraic genus~$\cg$ curve~$\Sigma$ and a system of~$\fA$,~$\fB$ cycles on it. Let $\eta_1,\dots,\eta_\cg$ be the basis of holomorphic differentials on~$\Sigma$ normalized by the condition $\oint_{\fA_i}\eta_j=\delta_{i,j}$ and $B$ is the Bergman kernel normalized on the $\fA$ cycles, that is $\oint_{\fA_i} B = 0$ and $\frac{1}{2\pi\ii}\int_{\fB_i} B = \eta_i$, $i=1,\dots,\cg$. The~$\cg\times \cg$ matrix~$\cT$ of $\fB$ periods $\cT_{i,j}=\oint_{\fB_i}\eta_j = \frac{1}{2\pi\ii}\oint_{\fB_i}\oint_{\fB_j} B$ is symmetric and~${\rm Im}\cT$ is positively definite. The \emph{Jacobian}~$J$ is defined as the quotient of~$\C^\cg$ by the $2\cg$-dimensional integer lattice
\begin{equation}
	J=\C^\cg/\{ m+\cT n\},\quad m, n\in\Z^\cg.
\end{equation}
The lattice is chosen so that the following \emph{Abel map} is well defined:
\begin{equation}
	\cA:\Sigma\to J,\quad p\mapsto \Bigl(\int_{q_0}^p\eta_1,\dots,\int_{q_0}^p\eta_\cg\Bigr)
\end{equation}
with the same contours for all integrals and for some base point~$q_0$ fixed in advance. The Abel map extends to divisors, $\cA(\sum n_i p_i)=\sum n_i\cA(p_i)$, or even to ${\rm Pic}\,\Sigma$, classes of linear equivalence of divisors: the values of~$\cA$ on equivalent divisors are the same.

The \emph{Riemann $\Theta$ function} is defined by an explicit formula
\begin{equation}
	\begin{aligned}
		\Theta&:\C^\cg\to\C,
		\\
		\Theta(w)&=\Theta(w|\cT)=\sum_{k\in\Z^\cg}e^{2\pi \ii ( \frac 12 k^T\cT k +k^Tw)}.
	\end{aligned}
\end{equation}
The convergence follows from the positive definiteness of~$\Im\cT$.
A shift of the argument~$w$ by a lattice vector does not preserve~$\Theta$ but multiplies it by an easily controlled factor. As a consequence, it does not define a function on the Jacobian but rather a section of certain line bundle over~$J$. In particular, for any vector~$w$, the ``function''  on~$\Sigma$ defined by
\begin{equation}
	F(p)=\Theta(\cA(p)+ w)
\end{equation}
is holomorphic (has no poles) but multivalued. Nevertheless, the divisor of its zeros (if~$F$ is not identically equal to zero) is well defined.

The function~$\Theta$ is even, $\Theta(w)=\Theta(-w)$. It is useful to consider also its odd analogue $\Theta_*$, $\Theta_*(-w)=-\Theta_*(w)$ defined as follows. Let $\mu=(\mu_1,\dots,\mu_\cg)$ and $\nu=(\nu_1,\dots,\nu_\cg)$ be two vectors (called \emph{theta-characteristics}) with the entries in~$\{0,\frac12\}$ such that the following congruence holds: $4\mu^T\nu\equiv 1\pmod2$. Then, we set
\begin{equation}
	\begin{aligned}
		\Theta_*(w)&=\Theta_{\left[\begin{smallmatrix}\mu\\\nu\end{smallmatrix}\right]}(w|\cT)
		=\sum_{ k\in\Z^\cg}e^{2\pi \ii\big( \frac 12  (k+\mu)^T\cT( k+\mu)+(k+\mu)^T(w+\nu)\big)}.
	\end{aligned}
\end{equation}

This function does depend on the choice of~$\mu$ and~$\nu$, but it is sufficient to fix one of these choices. In fact many relations also do not depend on this choice.
For example, we have the following expression for the Bergman kernel:
\begin{equation} \label{eq:B-in-terms-of-Theta}
	B(z_1,z_2)=d_1d_2\log\Theta_*(\cA(z_1-z_2)),
\end{equation}
independently of the freedom in the definition of~$\Theta_*$ \cite[Section IIIb, \S1]{Mumford}. 

\subsection{Prime form}\label{S2.2}
Let us expand~$\Theta_*$ at the origin,
\begin{equation}
	\Theta_*(w)=\theta_{*}w+(\text{higher order terms}),
\end{equation}
and use the coefficients of linear terms in this expansion to define the holomorphic differential
$d\zeta=\theta_{*}\eta$.
Also, introduce the bi-half-differential
\begin{equation} \label{eq:E-inverse}
	\frac{1}{E(z_1,z_2)}=\frac{\sqrt{d\zeta(z_1)}\sqrt{d\zeta(z_2)}}{\Theta_*(\cA(z_1-z_2))}.
\end{equation}
It is well defined in a neighborhood of the diagonal, but its global extension is multivalued. It should be considered as a section of a square root of the canonical bundle on $\Sigma$ in each of its variables, and the choice of the square root depends on the theta-characteristics.

Then, as we have already mentioned above in the Introduction, %
for any meromorphic function~$\chi$ that can serve as a local coordinate both at $z_1$ and $z_2$ on~$\Sigma$, the prime form \eqref{eq:E-inverse} is given by
\begin{equation}\label{eq:Eexp}
	\frac{1}{E(z_1,z_2)}=\frac{\sqrt{d\chi_1}\sqrt{d\chi_2}}{\chi_1-\chi_2}
	\exp\left({\frac12\int_{z_2}^{z_1}\int_{z_2}^{z_1}\bigl(B(\tilde z_1,\tilde z_2)-\frac{d\tilde \chi_1d\tilde \chi_2}{(\tilde \chi_1-\tilde \chi_2)^2}\bigr)}\right),
	\quad \chi_i=\chi(z_i),
\end{equation}
where $B$ is the Bergman kernel \eqref{eq:B-in-terms-of-Theta}.
The integrals on the right hand side are not defined without the regularization applied in them but it is easy to compute that the result is independent of the function~$\chi$ used for the regularization. This equality implies that $1/E(z_1,z_2)$ has no singularities outside the diagonal (but is still multivalued with possible branchings along non-contractible loops on~$\Sigma$). The function $\Theta_*(\cA(z_1-z_2))$ regarded as a function in $z_1$ has a zero at $z_2$ and $\cg-1$ more zeros, denoted by $r_1,\dots,r_{\cg-1}$. It follows that~$d\zeta$ has double zeros at $r_1,\dots,r_{\cg-1}$ and these points do not depend on $z_2$.

\begin{remark} \label{rem:PrimeFormE} Notice the following useful identity for the prime form $E(z_1,z_2)$ following from~\eqref{eq:Eexp} and
\begin{equation}
B(z_1,z_2)=d_1 d_2 \log (E(z_1,z_2))
\end{equation}
(which is equivalent to~\eqref{eq:B-in-terms-of-Theta} taking into account~\eqref{eq:E-inverse}),
 and appearing in some computations
\begin{align} \label{eq:exp-int-B-i-j}
	\exp\bigg(  \int_{z_1^-}^{z_1^+}\int_{z_2^-}^{z_2^+} B \bigg) = \frac{E(z_1^+,z_2^+)E(z_1^-,z_2^-)}{E(z_1^+,z_2^-)E(z_1^-,z_2^+)}.
\end{align}
\end{remark}

\subsection{Krichever differentials} \label{sec:Krichever-differentials} The Krichever construction associates a KP tau function to an arbitrary choice of a point $q_\infty\in\Sigma$ and a local coordinate $z$ at this point. It is given by an explicit formula
\begin{equation}\label{eq:Krtau}
\tau(t_1,t_2,\dots)=e^{\frac12\sum_{k,l\ge1}b_{k,l}t_kt_l}\frac{\Theta_*(\sum_{k=1}^\infty t_k U_k+w)}{\Theta_*(w)},
\end{equation}
where the vectors $U_k\in\C^\cg$ and the constants $b_{k,l}$ are determined by the following expansions at the point $q_\infty$:
\begin{align}
\label{eq:Uka}
\eta(z)&=\sum_{k=1}^\infty U_k z^{k-1}dz,
\\\label{eq:Bz12}
B(z_1,z_2)&=\frac{dz_1dz_2}{(z_1-z_2)^2}+\sum_{k,l\ge1}b_{k,l}z_1^{k-1}dz_1\;z_2^{l-1}dz_2.
\end{align}
Note that~\eqref{eq:Krtau} does depend on the choice of the theta-characteristics (and it is used in applications, cf. Corollary~\ref{cor:KP-deformed} below), but this choice does not affect the KP integrability properties. 

\begin{proposition}[\cite{Krichever-main}]
$\tau$ is a KP tau function for an arbitrary (generic) choice of the $\cg$-dimensional parameter~$w$.
\end{proposition}

The $n$-point differentials $\omega^\Kr_n$ associated with this tau function according to~\eqref{eq:n-differential} extend globally and are given explicitly (see \cite{Kawamoto}) by
	\begin{align}\label{eq:Kawamoto}
		\omega^\Kr_n = \left(\prod_{i=1}^n \eta(z_i) \partial_w \right) \log \Theta_*(w|\cT) + \delta_{n,2} B.
	\end{align}
Indeed, we see from~\eqref{eq:Krtau} that the loop insertion operator $\delta_z$ acts on $\log \tau$  by
\begin{equation}
\delta_z=\sum_{k=1}^\infty z^{k-1}dz\; \frac{\partial}{\partial t_k}=\sum_{k=1}^\infty z^{k-1}dz\; U_k \partial_w=\eta(z)\partial_w
\end{equation}
with a small correction implied by quadratic terms. This implies~\eqref{eq:Kawamoto}. Similar computations provide also an explicit expression for the $n|n$ half-differentials:
\begin{align} 
	\label{eq:Omegabnalg}
	&
	\Omega^{\Kr,\bullet}_n(z^+_{\set n},z^-_{\set n})=
	\\ \notag &
	\prod_{1\le i<j\le n}\frac{E(z^+_i,z^+_j)\,E(z^-_i,z^-_j)}{E(z^+_i,z^-_j)\,E( z^-_i,z^+_j)}\;
	\prod_{i=1}^n\frac{1}{E(z^+_i,z^-_i)}
	\frac{\Theta_*(\cA(\sum_{i=1}^n(z^+_i-z^-_i))+w)}{\Theta_*(w)}.
\end{align}

The KP integrability means that these differentials obey determinantal identities
\begin{align} \label{eq:Krichever-Fay}
	\omega^{\Kr}_n(z_{\set n}) &= \det\nolimits^\circ (\Omega^{\Kr}_1(z_{i},z_{j})  ),\quad n\ge2,	
\\	\Omega^{\Kr,\bullet}_n(z^+_{\set n},z^-_{\set n}) &= \det (\Omega^{\Kr}_1(z^+_{i},z^-_{j})  ).	
\end{align}	

The differential
\begin{equation}
\Omega^{\Kr}_1(z^+,z^-)=\frac{\Theta_*(\cA(z^+-z^-)+w)}{E(z^+,z^-)\Theta_*(w)}
\end{equation}
appearing on the right hand side is known as the Szeg\"o kernel.

\begin{remark} Equation~\eqref{eq:Krichever-Fay} for $n=2$ is a version of the so-called Fay trisecant identity.

For example, we have
\begin{align}
	\omega^\Kr_2(z_1,z_2)&=-\Omega^{\Kr}_1(z_1,z_2)\Omega^{\Kr}_1(z_2,z_1)
	\\\notag &=\frac{\Theta_*(\cA(z_1-z_2)+w)\,\Theta_*(\cA(z_2-z_1)+w)}{E(z_1,z_2)^2\Theta_*(w)^2}.
\end{align}
\end{remark}

\begin{remark}
Note that we get not just a single KP integrable system of differentials but rather a family of those parameterized by~$w$ and satisfying
\begin{equation}
\omega^\Kr_{n+1}(z,z_{\set n})-\delta_{n,1}B(z,z_1)=\eta(z)\,\partial_w\, \big(\omega^\Kr_{n}(z_{\set n}) -\delta_{n,2} B(z_{\set 2}) \big).
\end{equation}
Also note that the freedom in a choice of the theta characteristics entering the definition of $\Theta_*$ does not affect the KP integrability: a different choice $\Theta_*$ leads to a shift of~$w$, multiplication of $\Theta_*$ by a constant, and by a factor whose logarithm is linear in~$w$. All these transformations are obvious KP symmetries  (recall that the KP equations have constant coefficients and are at least quadratic in derivatives). For example, the last one affects the differential $\omega^{\Kr}_1$ only.
\end{remark}

\section{Basic constructions and properties}

The proofs of the statements in this section are relegated to Section~\ref{sec:Proofs}.

\label{sec:constr-and-prop}

\subsection{Definitions of non-perturbative objects} \label{sec:np-definitions}
Let $(\Sigma,dx,dy,\cP)$ be the initial data of (generalized) topological recursion, with the standard Bergman kernel,
and arrange the differentials that it produces into the series
\begin{align}
	\omega_m \coloneqq \sum_{\scriptstyle \substack{g\geq 0 \\ 2g-2+m >0}} \hbar^{2g-2+m} \omega^{(g)}_m.
\end{align}
Here and below the expansion in $\hbar$ is always formal, and the formulas and results as presented are valid in the realm of formal power series expansions in $\hbar$.

\begin{definition}[\cite{BorEyn-AllOrderConjecture}]\label{def:np-differentials} The non-perturbative disconnected $n|n$ half-differentials $\Omega_n^{\mathsf{np},\bullet}$ are defined as
\begin{align} \label{eq:np-nn-kernels} 
	& \Omega_n^{\np,\bullet}(z^+_{\set{n}},z^-_{\set n})   \coloneqq 
	\\ \notag & 
	\prod_{1\leq k<l\leq n} \frac{E(z^+_k,z^+_l)E(z^-_k,  z^-_l)}{E(z^+_k, z^-_l)E(z^-_k,z^+_l)}  \prod_{i=1}^n \frac{\exp \Big(\frac 1\hbar  \int_{z^-_i}^{z^+_i} \omega^{(0)}_1 \Big)}{E(z^+_i, z^-_i)}  \times
	\\ \notag & 
	 \left(
		{
			\exp\Bigg(\sum_{\scriptstyle m\geq 1}\frac{1}{m!} \Big(\frac{1}{2\pi\ii}\oint_{\fB} {\partial_w}\Big)^m  \omega_m\Bigg) \Theta_*(w|\cT)
		} \right)^{-1}\times
	\\ \notag &  
	\exp\left(
	\sum_{\scriptstyle\substack{m_0,\dots,m_n\geq 0 \\ \sum_{i=0}^n m_i=m\geq 1}}  \frac{1}{\prod_{i=0}^n m_i!} \Big(\frac{1}{2\pi\ii}\oint_{\fB} {\partial_w}\Big)^{m_0} \prod_{i=1}^n \Big(\int_{z^-_i}^{z^+_i} \Big)^{m_i} \omega_m
	\right)
	\\ \notag & 
	\Theta_*\big(w+\cA\Big(\sum_{i=1}^n (z^+_i- z^-_i)\Big)\big|\cT\big)
\end{align}
\end{definition}

This formula contains infinite summation and makes sense only in expansion in $\hbar$. Note that as presented, the explicit dependence on $\hbar$ is contained only in the exponential prefactor and in $\omega_m$, and there is no explicit dependence on $\hbar$ in the $\Theta_*$ factors. However, in non-perturbative applications this might be distorted, as $w$ might be shifted by an $\hbar$-dependent term. The internal structure of this expression is described in the following Lemmas.

\begin{lemma}\label{lem:omega-np-def} The differentials $\Omega^{\np,\bullet}_n$ are extended $n|n$ disconnected half-differentials associated with a certain family of symmetric $n$-differentials $\omega^\np_n$ by the following slight variation of formulas of Section~\ref{sec:KP}:

\begin{align} \label{eq:incl-excl-np}
	\Omega_n^{\np}(z^+_{\set{n}},z^-_{\set n})  &\coloneqq \sum_{\ell=1}^n \frac{(-1)^{\ell-1}}{\ell} \sum_{\scriptstyle\substack{I_1\sqcup \cdots \sqcup I_\ell = \set n \\ \forall j I_j\not=\emptyset}} \prod_{j=1}^\ell \Omega_{|I_j|}^{\np,\bullet}(z^+_{I_j},z^-_{I_j}),
	\\
	\bW^\np_n(z_{\set n}, u_{\set n}) &\coloneqq
	\Big(\prod_{i=1}^n
	e^{-u_i y_i}\Big)
	\restr{z_i^\pm}{e^{\pm \frac{u_i\hbar}{2}\partial_{x_i}} z_i}
	\Omega_n^{\mathsf{np}}(z^+_{\set{n}},z^-_{\set n}),\label{Wom}
	\\	
	\label{eq:cW-np}
	\cW^\np_n(z,u;z_{\set n}) & \coloneqq \bW^\np_{n+1}(z,z_{\set n}, u, u_{\set n})\big\vert_{u_{\set n} = 0},
	\\
	\omega^\np_n (z_{\set n}) & =\restr{u_{\set n}}{0} \left(\bW^\np_n(z_{\set n}, u_{\set n})-\delta_{n,1}\frac{dx_1}{u_1\hbar}\right),
\end{align}
and $\Omega^{\np,\bullet}_n$, $n\geq 1$, can be reconstructed from $\omega^\np_n$, $n\geq 1$, as
\begin{align}
	\label{eq:Omeganp-intermsof-omeganp}
	& \Omega^{\np,\bullet}_n(z^+_{\set n},z^-_{\set n})  = 	
	\\ \notag &
	 \prod_{1\leq k<l\leq n} \frac{E(z^+_k,z^+_l)E(z^-_k,  z^-_l)}{E(z^+_k, z^-_l)E(z^-_k,z^+_l)}
	\prod_{i=1}^n \frac{\exp \Big( \frac 1\hbar \int_{z^-_i}^{z^+_i} \omega^{(0)}_1 \Big)
	}{E(z^+_i, z^-_i)} \times 
		\\ \notag
	& 
	\exp\Bigg(
	\sum_{\scriptstyle\substack{m_1,\dots,m_n\geq 0 \\ \sum_{i=1}^n m_i=m\geq 1}}
	\prod\limits_{i=1}^n \frac{1}{m_i!} \Big(\int_{z_i^-}^{z_i^+} \Big)^{m_i} (\omega^\np_m-\delta_{m,2}B)
	\Bigg).	
\end{align}
The corresponding non-perturbative n-differentials $\omega^\np_n$ are formal power series in $\hbar$, whose coefficients are finite algebraic expressions in terms of the integrals of $\omega^{(g)}_m$ and derivatives of $\log \Theta_*(w|\cT)$.
\end{lemma}

\begin{remark} Note that the contribution of $\omega^{(0)}_1=y\,dx$ is included in \eqref{eq:np-nn-kernels}, for the reason of consistency with the definitions in the other sources. This contribution does not affect the KP integrability but might be useful, for instance, for
	presentation of the loop equations. However, these terms are not included to~$\omega_n$ or $\omega^\np_n$, according to our convention, in order to avoid appearance of negative powers of~$\hbar$. By that reason, the relation between $\Omega^\np_n$ and $\bW^\np_n$ and $\cW^\np_n$ is slightly different from those of~\eqref{eq:bW-general}--\eqref{eq:cW-general}.
\end{remark}

\begin{lemma}\label{lem:omega-np-analytic} The leading term  of the expansion of $\omega^\np_n$ in $\hbar$ is the corresponding Krichever differential,
\begin{equation}\label{eq:omeganpomegakr}
\omega_n^\np=\omega^\Kr_n+O(\hbar).
\end{equation}
In particular, $\restr\hbar0\omega^\np_n-\delta_{n,2}B=\prod_{j=1}^n\eta(z_j)\partial_w\log\Theta_*(w|\cT)$ is holomorphic and $\omega_n^\np=\omega^\Kr_n$ if $\omega^{(g)}_m=0$, $2g-2+m>0$.

The coefficient of any positive power of $\hbar$ in~$\omega^\np_n$ is a global symmetric meromorphic differential with the only poles in each variable at the points $q\in\cP$.
\end{lemma}

\begin{remark} Note that the expansion of the non-perturbative $n$-differentials in $\hbar$ is not a topological expansion. But we still can define $\omega^{\np,\langle d\rangle}_n (z_{\set n}) \coloneqq [\hbar^d] \omega^\np_n (z_{\set n})$, $d\geq 0$. An existence of topological expansion would mean that these components could be nonzero only for the values of $d$ of the form $2g-2+n$, $g\in\Z_{\ge0}$. But we have, in general, contributions for all nonnegative integers~$d$.
\end{remark}

\begin{remark}
	Explicitly, $\omega^\np_n$ is defined in terms of $\{\omega_m\}$ and $\log\Theta_*$ by a closed combinatorial formula with summation over graphs. As a corollary, we obtain similar combinatorial formulas for all versions of extended differentials $\Omega^{\np,\bullet}_n$, $\Omega^\np_n$, $\bW^\np_n$, $\cW^\np_n$ associated to $\{\omega^\np_n\}$ by Equations~\eqref{eq:Omeganp-intermsof-omeganp}, \eqref{eq:incl-excl-np}--\eqref{eq:cW-np}. All these combinatorial expressions are reviewed in Sect.~\ref{sec:Proofs}.
\end{remark}

\subsection{Loop equations and projection property}
Assume the standard setup of topological recursion.
Define
\begin{align}
	\cW^{\np,\langle d\rangle}_n & \coloneqq [\hbar ^d] \cW^\np_n(z,u;z_{\set n}).
\end{align}

\begin{theorem}\label{thm:loop-equations} The non-perturbative differentials satisfy the loop equations. Namely, for any \xvital{}-point $q\in\cP$ %
and for any $k\ge0$  we have:
\begin{equation} \label{eq:(k-1)-loop-eq}
	[u^k]e^{u y}\cW^{\np,\langle d\rangle}_n(z,u;z_{\set n})\in\Xi_q,
\end{equation}
where $y$ is any local primitive of $dy$.  This relation holds identically in $z_1,\dots,z_n$.
\end{theorem}

\begin{remark} As in the usual perturbative case~\cite{ABDKS1,alexandrov2024degenerateirregulartopologicalrecursion}, an equivalent way to state the loop equations is to demand that
\begin{equation}\label{eq:mainTRrelation}
	\sum_{r\ge0}\Bigl(-d\frac{1}{dy}\Bigr)^r[u^r]\cW^{\np,\langle d \rangle}_{n}(z,u;z_{\set {n}}) \text{ is holomorphic at } z=q \text{ for } q\in\cP.
\end{equation}
Moreover, this relation holds as well in the general setting, with arbitrary orders of zeros/poles of~$dx$ and~$dy$ at the $\xvital$-points. %
\end{remark}

Theorem~\ref{thm:loop-equations} allows to compute the principal parts of $\omega^{\np,\langle d\rangle}_n$ in each variable once we know $\omega^{\np,\langle d'\rangle}_{n'}$ with $d'+n'<d+n$.
These principal parts allow us to reconstruct  $\omega^{\np,\langle d\rangle}_n$ for $d\geq 1$, $n\geq 1$, uniquely from $\omega^{\np,\langle 0\rangle}_n$, $n\geq 1$, using the following theorem

\begin{theorem} \label{thm:projectionproperty} For any $d\geq 0$, $n\geq 0$, $1\leq i \leq \cg$ we have the following projection property:
\begin{align} \label{eq:loop-deformation}
	\oint_{z\in \fA_i} \big(\omega^{\np,\langle d\rangle}_{n+1}(z,z_{\set n})
	-\eta(z)\partial_w  \omega^{\np,\langle d\rangle}_{n}(z_{\set n})\big) = 0,
\end{align}	
where for $n=0$
\begin{equation}\label{eq:omeganp0}	
\omega^{\np,\langle d\rangle}_0\coloneqq [\hbar^d] \log \left( \exp\left(\sum_{m\geq 1}\frac{1}{m!} \Big(\frac{1}{2\pi\ii}\oint_{\fB} {\partial_w}\Big)^m  \omega_m\right) \Theta_*(w|\cT) \right).
\end{equation}
\end{theorem}

\begin{remark} This setup is very close to the setup
	of the so-called blobbed topological recursion~\cite{BS-blobbed}. With this analogy the role of blobs in the present construction is played by $\big( \prod_{i=1}^n \eta(z_i) \partial_w\big) \log \Theta_*(w|\cT)$. Note, however, that the expansion in $\hbar$ is non-topological. The coupling to blobs might also look a bit different at the first glance, but geometry of the underlying Riemann surface allows to reduce the $\oint_\fB$ integrals to the residues at the points in $\cP$.

Let us mention that this setup can also be applied to generalized topological recursion, but it requires to develop a relevant piece of theory, see Section~\ref{sec:Blobs-Gen}  and~\cite{alexandrov2025blobbedtopologicalrecursionkp}.
\end{remark}

\begin{remark} Note also that Equation~\eqref{eq:loop-deformation} has also an alternative interpretation as a deformation formula. Namely, we can rewrite it as
	\begin{align} \label{eq:loop-deformation-2}
		 \partial_{w_i}  \omega^{\np,\langle d\rangle}_{n}(z_{\set n}) =\oint_{z\in \fA_i} \omega^{\np,\langle d\rangle}_{n+1}(z,z_{\set n}).
	\end{align} 
\end{remark}

\subsection{The \texorpdfstring{$x-y$}{x-y} swap relation and \texorpdfstring{$dy$}{dy} deformations} Assume the generalized setup of topological recursion.

Consider the dual topological recursion with the initial data $(\Sigma,dy,dx,\cP^\vee)$. We use the differentials $\omega^{\vee,(g)}_n$, that it produces, to construct exactly the same sequence of non-perturbative objects, that is, $\Omega_n^{\np,\vee,\bullet}$, $\Omega_n^{\np,\vee}$, $\bW_n^{\np,\vee}$, $\omega_n^{\np,\vee}$, and $\cW_n^{\np,\vee}$, defined for exactly the same theta-characteristics and parameters $w$. Then the $x-y$ swap formula can be extended to the non-perturbative context:

\begin{theorem} \label{thm:xyswap}
The systems of non-perturbative multi-differentials for $(\Sigma,dx,dy,\cP) $ and $(\Sigma,dy,dx,\cP^\vee)$ are related by the following formulas (as formal expansions in $\hbar$):		
		\begin{align}\label{eq:xyrel}
			\restr{w}{-w}\omega^{\np,\vee}_n(z_{\set n}) & =(-1)^n
			\left(\prod_{i=1}^n\sum_{r=0}^\infty \Bigl(-d_i\frac{1}{dy_i}\Bigr)^{r}[u_i^r]\right)
			\bW^{\np}_n(z_{\set n},u_{\set n}); \\
			\restr{w}{-w}\omega^\np_n(z_{\set n}) & =(-1)^n
			\left(\prod_{i=1}^n\sum_{r=0}^\infty \Bigl(-d_i\frac{1}{dx_i}\Bigr)^{r}[u_i^r]\right)
			\bW^{\np,\vee}_n(z_{\set n},u_{\set n}).\label{eq:xyrel1}
		\end{align}
\end{theorem}

 A useful tool in the theory of topological recursion is the effect of the deformations of the initial data $dx$ and $dy$. Our goal is to find the corresponding formulas in the non-perturbative case. Since we can use the $x-y$ swap formula discussed above, it is sufficient to study the deformations of $y$. However, in this case we get back to the standard setup of topological recursion in order to formulate the following.

Consider a deformation of $dy$ given by $dy+\epsilon d(\Delta y)$ for small $\epsilon$, where $\Delta y$ is some local primitive for the deformation of $dy$ near the points in $\cP$. Assume that $d(\Delta y)$ is regular at the points in $\cP$.  Let $\omega_n^{\np} + \epsilon \Delta \omega_n^{\np}$ be the corresponding deformations of the non-perturbative $n$-differentials.

\begin{theorem} \label{thm:Deformation-y} We have
	\begin{align} \label{eq:Deformation-y}
		\Delta \omega_n^{\np} (z_{\set n}) = \sum_{q\in\cP} \res_{z=q} \omega_{n+1}^{\np} (z,z_{\set n})\int_q^z \Delta y\, dx.
	\end{align}
\end{theorem}

\subsection{KP integrability} Assume the generalized setup of topological recursion.

The system of non-perturbative $n|n$ half-differentials and (extended) differentials satisfies the following KP integrability property expressed as determinantal formulas, which is the main theorem of the present paper.

\begin{theorem}\label{thm:KP-integrable} The system of differentials $\{\omega^{\np}_n\}$ is KP integrable, that is, we have:
	\begin{align} \label{eq:KP-integrable-smallomega}
		\omega^{\np}_n (z_{\set n}) & = {\det}^\circ (\Omega^\np_1 (z_i,z_{j})),\quad n\ge2,
\\		\Omega^{\np,\bullet}_n (z^+_{\set n},z^-_{\set n}) & = {\det} (\Omega^\np_1 (z^+_i,z^-_{j})).
	\end{align}
\end{theorem}
The proof of Theorem~\ref{thm:KP-integrable} is given in Section~\ref{sec:proofKP}.

In order to incorporate slight changes in the choice of conventions in the definitions of the non-perturbative objects in~\cite{EynardMarino,BorEyn-AllOrderConjecture,BorEyn-knots,eynard2024hirotafaygeometry}, we also state the following slight variation of Theorem~\ref{thm:KP-integrable}:

\begin{corollary} \label{cor:KP-deformed} Let $\{\widetilde\omega^{\np}_n\}$ be a system of non-perturbative differentials defined by the same formulas as in Definition~\ref{def:np-differentials}, but for a different choice of theta characteristics entering the definition of $\Theta_*$, or even where we replace $\Theta_*(w|\cT)$ in Equation~\eqref{eq:np-nn-kernels} by
$\widetilde\Theta_*(w|\cT)\coloneqq ae^{b^Tw} \Theta_*(w|\cT)$ 
for an arbitrary choice of constant $a$ and vector $b$. Moreover, in the final expressions for $\{\widetilde\omega^{\np}_n\}$ we can choose arbitrary values of the parameters $w$.
Then $\{\widetilde\omega^{\np}_n\}$ is KP integrable.
\end{corollary}

Indeed, this modification of $\Theta_*$ leads to a shift of $w$ or to a modification of $\omega^\Kr_1$ in the Krichever construction, which obviously preserves KP integrability of Krichever differentials and the whole proof of Theorem~\ref{thm:KP-integrable} works for a modified~$\Theta_*$, with an extra caveat that concerns the change of the argument of $\Theta_*$ under $x-y$ swap.

\begin{remark}
For instance, in~\cite{EynardMarino} the authors use the following choice of parameters
\begin{align}
	a & = e^{\frac{1}{4\pi\ii \hbar^2} \left(\int_{\fA}\omega^{(0)}_1\right)^T \cT \int_{\fA}\omega^{(0)}_1-2\pi\ii \,  \mu^T\nu},   &
	b & =-\frac{1}{\hbar}\oint_{\cT\fA}\omega^{(0)}_1, &
\end{align}
and in the final expressions put
\begin{equation}
w  =\frac{1}{2\pi\ii \hbar}\oint_{\fB-\cT\fA}\omega^{(0)}_1
\end{equation}
in order to achieve the so-called background independence. Note that this leads to modified theta functions, and, subsequently, tau functions becoming trans-series in $\hbar$.
\end{remark}

\begin{remark}
Corollary~\ref{cor:KP-deformed} is also a tool to re-introduce the (multi-)KP times as it is done in~\cite{BorEyn-AllOrderConjecture,eynard2024hirotafaygeometry,krichever2023quasiperiodicsolutionsuniversalhierarchy}, which we essentially suppressed everywhere so far for the clarity of exposition except for Equation~\eqref{eq:Krtau}. To this end, we select a number of points $p_1,\dots,p_N\in\Sigma\setminus \cP$, $N\geq 1$, and consider the choices of the second kind differentials $d\Upsilon^{\alpha}_{k}$ with the only pole at $p_\alpha$ of order $k+1$, $\alpha=1,\dots,N$, normalized by the condition $\oint_\fA d \Upsilon^{\alpha}_{k} = 0$. Then the following choice of parameters introduces the multi-KP variables $t^\alpha_k$, $\alpha=1,\dots,N$, $k\geq 1$:
\begin{align}
	a & = e^{\frac 12 \sum_{\alpha,k}\sum_{\beta,l} \Upsilon^{\alpha\beta}_{kl} t^{\alpha}_k t^\beta_l},   &
	b & =0, &
	w & =\frac{1}{2\pi\ii}\oint_{\fB} \sum_{\alpha,k}t^\alpha_k d\Upsilon^{\alpha}_k,
\end{align}
where $\Upsilon^{\alpha\beta}_{kl}$ is a bilinear form obtained by expanding the holomorphic parts of the primitives of $d\Upsilon^{\alpha}_{k}$ at $p_\beta$. We refer to~\cite[Sections 3, 4, 5]{krichever2023quasiperiodicsolutionsuniversalhierarchy} for a detailed exposition; note that this construction can be enhanced to include more variables of ``discrete'' kind, cf. also a review in~\cite[Examples 2 and 3]{KricheverShiota}. We further discuss the KP case in Section~\ref{sec:discussion-formulations}.	
\end{remark}
\begin{remark}
Krichever's differentials $\Omega^{\Kr,\bullet}_n$ can be interpreted as free fermion correlation functions of the Riemann surface, see e.g. \cite{Ooguri}. Then the non-perturbative differentials $\Omega^{\np,\bullet}_n$ correspond to insertion of Zamolodchikov's operators \cite{Zamo} in the standard setup, or their generalizations in the case of generalized topological recursion.
\end{remark}

\section{Proofs}\label{sec:Proofs}

\subsection{Graphical formulas} The proofs are based on the graphical formula for $\omega^{\np}_n$ proposed in~\cite{BorEyn-knots}. More precisely, in~\cite{BorEyn-knots} the authors give graphical formulas for the non-perturbative connected $n|n$ half-differentials $\Omega^\np_n$, and we further specialize them here in order to deal with $\omega^\np_n$ and $\cW^\np_n$.

In the rest of this section, as well as in the next section, we use a variety of (different!) graphs that share a lot of common features. In each formula below we define a set $G_\star$ of 3-level graphs $\Gamma$ with some additional structure, and the common grounds are the following:

\begin{itemize}
	\item The set of vertices $V(\Gamma)$ splits into three subsets $V(\Gamma)=V_\ell(\Gamma)\sqcup V_\omega(\Gamma)\sqcup V_{\theta}(\Gamma)$, which we call (multi)leaves, $\omega$-vertices and $\theta$-vertices, respectively.
	\item Each edge in the set of edges $E(\Gamma)$ connects an $\omega$-vertex to either a multileaf or a $\theta$-vertex. Accordingly, the set $E(\Gamma)$ splits as $E(\Gamma)=E_{\ell-\omega}(\Gamma)\sqcup E_{\omega-\theta}(\Gamma)$.
	\item Graph $\Gamma$ must be connected.
	\item The set $V_\ell(\Gamma)$ consists of $n$ or, in some cases, of $n+1$ ordered multileaves, depending on a particular problem that we address.
	\item 	Each edge $e\in E_{\ell-\omega}(\Gamma)$ is decorated by an operator $O(e)$ acting on the decoration of the $\omega$-vertex where $e$ is attached. The operators have to be specified in each case.
	\item There is a map $g\colon V_\omega(\Gamma)\to \Z_{\geq 0}$.
	An $\omega$-vertex $v$ of $\Gamma$ with $m(v)$ edges in $E_{\ell-\omega}(\Gamma)$ and $k(v)$ edges in $E_{\omega-\theta}(\Gamma)$ attached to it and labeled by $g(v)$ is called
	a $(g(v),m(v),k(v))$-$\omega$-vertex.
	\item We require that $2g(v)-2+m(v)+k(v)\geq 0$ for each $\omega$-vertex $v$.
	\item The $(0,0,2)$-$\omega$-vertices  are not allowed.
	\item The two edges attached to the same $(0,2,0)$-$\omega$-vertex cannot be also attached to the same leaf.
	\item The $(g,m,k)$-$\omega$-vertices $v$ are decorated by some differentials $\omega(v)$. These decorations have to be specified depending on a particular problem that we address.
	\item
	Each $\theta$-vertex $v$ is decorated by $\theta(v)\coloneqq \log\Theta_*(w|\cT)$.
	\item Each edge $e\in E_{\omega-\theta}(\Gamma)$ is decorated by the bi-linear operator  \begin{align} 
		O(e)\coloneqq \frac{1}{2\pi\ii}\int_\fB \partial_w,
	\end{align}
 	where the $\int_{\fB_i}$ operators act on the decoration of the attached $\omega$-vertex and $\partial_{w_i}$ act on the decoration of the attached $\theta$-vertex. 
	\item $|\Aut(\Gamma)|$ denotes the order of the automorphisms group of a decorated graph $\Gamma$ that preserves all decorations.
	\item We associate to a graph $\Gamma$ its weight $\mathsf{w}_\star (\Gamma)$ defined as
	\begin{align}
		\mathsf{w}_\star(\Gamma) \coloneqq \frac{1}{|\Aut(\Gamma)|}  \prod_{e\in E(\Gamma)} O(e)
		\bigg(\prod_{v\in V_\omega(\Gamma)} \omega(v) \prod_{v\in V_{\theta}(\Gamma)} \theta(v)\bigg),
	\end{align}
	where the action of the operators associated to the edges on the decorations of the vertices is prescribed by the graph $\Gamma$.
\end{itemize}

This structure is specified in each particular case further by the choice of the labels of (multi)leaves and specific conditions on the index of these vertices (sometimes that are indeed of index $1$, hence the name ``leaves''), associated operators $O(e)$ for $e\in E_{\ell-\omega}(\Gamma)$, and the decorations $\omega(v)$ for the $(g,m,k)$-$\omega$-vertices.

\subsubsection{Borot--Eynard formula}
Let $G_{\Omega_n}$ be the set of 3-level graphs $\Gamma$ with some additional structure described as follows:
\begin{itemize}
	\item The set $V_\ell(\Gamma)$ consists of $n$ ordered multileaves $\ell_1,\dots,\ell_n$, that is, there is a fixed bijection $\set n \to V_\ell(\Gamma)$ such that $i\mapsto \ell_i$.
	\item 	Each edge $e\in E_{\ell-\omega}(\Gamma)$ attached to $\ell_i$ is decorated by the operator $O(e)\coloneqq \int_{z^-_i}^{z^+_i}$.
	\item Each $(g,m,k)$-$\omega$-vertex $v$ is decorated by $\omega(v)\coloneqq \hbar^{2g-2+m+k}\omega^{(g)}_{m+k}$, and the arguments of $\omega^{(g)}_{m+k}$ are put in bijection with the edges attached to $v$.
\end{itemize}

\begin{proposition}[\cite{BorEyn-knots}]
\label{prop:Graphs-Omega}	
We have
\begin{align} \label{eq:GraphicalFormulaOmega}
	\Omega^\np_n =  \prod_{i=1}^n \frac{\exp \bigg( \frac 1\hbar \int_{z^-_i}^{z^+_i} \omega^{(0)}_1 
		\bigg)}{E(z^+_i,z^-_i)} \sum_{\Gamma\in G_{\Omega_n}} \mathsf{w}_{\Omega_n}(\Gamma).
\end{align}	
\end{proposition}

\begin{remark}\label{rem:Graphs-02} Proposition~\ref{prop:Graphs-Omega} is already presented in \cite{BorEyn-knots}; we just formalize their description in a way that is convenient for our proofs below. To this end, in addition to the standard Feynman graphs technique one has to use two additional observations. First of all, we use Equation~\eqref{eq:exp-int-B-i-j} in order to dissolve the first factor in~\eqref{eq:np-nn-kernels} into the $(0,2,0)$-$\omega$-vertices connected to two different multileaves. Moreover, we use the identity
\begin{align} \label{eq:theta-shift-Krichever}
		\Theta_*\big(w+\cA\bigg(\sum_{i=1}^n (z^+_i- z^-_i)\bigg)\big|\cT\big) = \exp\bigg(\sum_{i=1}^n \frac{1}{2\pi\ii}\int_\fB\int_{z^-_i}^{z_i^+} B \partial_w  \bigg) \exp ( \log \Theta_*(w|\cT))
\end{align}
to generate the $(0,1,1)$-$\omega$-vertices whose two edges are connected to a multileaf and to a $\theta$-vertex.
\end{remark}

\begin{remark} The right hand side of~\eqref{eq:GraphicalFormulaOmega} contains an infinite sum. However, both the left hand side and the right hand side of~\eqref{eq:GraphicalFormulaOmega} are formal power series expansions in $\hbar$. Then for each $d$ there is still an infinite number of graphs $\Gamma\in G_{\Omega_n}$ that contribute non-trivially to the coefficient of $\hbar^d$ in the $\hbar$-expansion, but once we remove all $(0,1,1)$ vertices we get just a finite number of possible ``core'' graphs. Since the contribution of all possible configurations with $(0,1,1)$-vertices is of exponential type, see Remark~\ref{rem:Graphs-02}, there is no problem with convergence.
\end{remark}

Proposition~\ref{prop:Graphs-Omega} can easily be specialized to the definitions of $\omega^\np_n$ and $\cW^\np_n$, let us describe the necessary modifications.

\subsubsection{A formula for $\omega_n^\np$}
Let $G_{\omega_n}$ be the set of 3-level graphs $\Gamma$ with some additional structure described as follows:
\begin{itemize}
	\item The set $V_\ell(\Gamma)$ consists of $n$ ordered leaves $\ell_1,\dots,\ell_n$, that is, there is a fixed bijection $\set n \to V_\ell(\Gamma)$ such that $i\mapsto \ell_i$.
	\item There is exactly one edge attached to each $\ell_i$.
	\item 	The edge $e\in E_{\ell-\omega}(\Gamma)$ attached to $\ell_i$ is decorated by the operator $O(e)\coloneqq \restr {} {z_i}$.
	\item Each $(g,m,k)$-$\omega$-vertex $v$ is decorated by $\omega(v)\coloneqq \hbar^{2g-2+m+k}\omega^{(g)}_{m+k}$, and the arguments of $\omega^{(g)}_{m+k}$ are put in bijection with the edges attached to $v$.
\end{itemize}

A direct corollary of Proposition~\ref{prop:Graphs-Omega} is the following

\begin{corollary} \label{cor:Graphs-omega}
We have
\begin{align} \label{eq:GraphicalFormulaomega}
	\omega^\np_n  =  \sum_{\Gamma\in G_{\omega_n}} \mathsf{w}_{\omega_n}(\Gamma).
\end{align}	
\end{corollary}

\begin{remark} Equation~\eqref{eq:GraphicalFormulaomega} is an equality of two formal power series in $\hbar$. For each $d$ there is only a finite number of decorated graphs $\Gamma\in G_{\omega_n}$ that contribute non-trivially to the coefficient of $\hbar^d$ in the $\hbar$-expansion.
\end{remark}

\begin{remark} Note that the constant term in $\hbar$ in Equation~\eqref{eq:GraphicalFormulaomega} is indeed the one prescribed by Equation~\eqref{eq:omeganpomegakr}.
\end{remark}

\begin{remark} Note also that feeding Equation~\eqref{eq:GraphicalFormulaomega} into the right hand side of Equation~\eqref{eq:Omeganp-intermsof-omeganp}, and subsequently applying Equation~\eqref{eq:incl-excl-np}, we get back to Equation~\eqref{eq:GraphicalFormulaOmega} for $\Omega_n^\np$.
\end{remark}

\begin{remark}\label{rem:om0taut}
	Finally, note that~\eqref{eq:GraphicalFormulaomega} makes sense also for $n=0$, where the left hand side is defined by~\eqref{eq:omeganp0}, and then it holds tautologically.
\end{remark}

\subsubsection{A formula for $\cW_n^\np$}
Let $G_{\cW_n}$ be the set of 3-level graphs $\Gamma$ with some additional structure described as follows:
\begin{itemize}
	\item The set $V_\ell(\Gamma)$ consists of $n+1$ ordered (multi)leaves $\ell_0,\ell_1,\dots,\ell_n$, that is, there is a fixed bijection $\{0\}\cup \set n \to V_\ell(\Gamma)$ such that $i\mapsto \ell_i$.
	\item There is exactly one edge attached to each $\ell_i$ for $i\in \set n$.
	\item 	The edge $e\in E_{\ell-\omega}(\Gamma)$ attached to $\ell_i$, $i\in \set n$,  is decorated by the operator $O(e)\coloneqq \restr {}{z_i}$.
	\item Each edge $e\in E_{\ell-\omega}(\Gamma)$ attached to $\ell_0$  is decorated by the operator $O(e)\coloneqq u\hbar \cS(u\hbar \partial_{x(z)})\frac{1}{dx(z)}\restr{}{z}$.
	\item Each $(g,m,k)$-$\omega$-vertex $v$ is decorated by $\omega(v)\coloneqq \hbar^{2g-2+m+k}\omega^{(g)}_{m+k}$, and the arguments of $\omega^{(g)}_m$ are put in bijection with the edges attached to $v$.
\end{itemize}

A direct corollary of Proposition~\ref{prop:Graphs-Omega} is the following

\begin{corollary} \label{cor:Graphs-cW}
	We have
	\begin{align} \label{eq:GraphicalFormula-cW}
		\cW^\np_n & =
		\frac{dx}{u\hbar} \exp\bigg({\restr{z'}{z}\restr{z''}{z}u\hbar\cS(u\hbar \partial_{x'})u\hbar\cS(u\hbar \partial_{x''}) \Big(\frac{B(z',z'')}{dx'dx''} - \frac{1}{(x'-x'')^2}\Big)} \bigg) 
		 \\ \notag & \quad
		 		 \times \exp\big({u\cS(u\hbar \partial_x) y - uy}\big) \sum_{\Gamma\in G_{\cW_n}} \mathsf{w}_{\cW_n}(\Gamma).
	\end{align}	
	Here $x=x(z),x'=x(z'),x''=x(z'')$.
\end{corollary}

\subsection{Proof of the loop equations} In this Section we prove Theorem~\ref{thm:loop-equations}. To this end, we use Corollary~\ref{cor:Graphs-cW}. In the corresponding graph formula we see that we can recollect all prefactors in~\eqref{eq:GraphicalFormula-cW} and the weights of all $\omega$-vertices attached to $\ell_0$ and the operators on the edges incident to $\ell_0$ into known graphical formulas for $\cW_m$ for some $m\geq 0$. Thus, we have the following lemma:

\begin{lemma} \label{lem:resummation} We have
	\begin{align} \label{eq:NewGraphicalFormula-cW}
		\cW^\np_n =
	\sum_{\Gamma\in G_{\cW'_n}}  {\mathsf{w}}_{\cW'_n}(\Gamma),
	\end{align}	
where $G_{\cW'_n}$ is the set of 3-level graphs $\Gamma$ with some additional structure described as follows:
\begin{itemize}
	\item The set $V_\ell(\Gamma)$ consists of $n+1$ ordered leaves $\ell_0,\ell_1,\dots,\ell_n$, that is, there is a fixed bijection $\{0\}\cup \set n \to V_\ell(\Gamma)$ such that $i\mapsto \ell_i$.
	\item There is exactly one edge attached to each $\ell_i$ for $i\in \{0\}\cup \set n$.
	\item 	The edge $e\in E_{\ell-\omega}(\Gamma)$ attached to $\ell_i$, $i\in \set n$,  is decorated by the operator $O(e)\coloneqq \restr{} {z_i}$.
	\item The edge $e\in E_{\ell-\omega}(\Gamma)$ attached to $\ell_0$  is decorated by the operator $O(e)\coloneqq \restr{} {z}$.
	\item Each $(g,m,k)$-$\omega$-vertex $v$ not connected to $\ell_0$ is decorated by \begin{align}
		\omega(v)\coloneqq \hbar^{2g-2+m+k}\omega^{(g)}_{m+k},
	\end{align} and the arguments of $\omega^{(g)}_m$ are put in bijection with the edges attached to $v$. 
	\item The only $(g,m+1,k)$-$\omega$-vertex $v$ that is connected to $\ell_0$ is decorated by $\omega(v)\coloneqq \hbar^{2g-1+m+k}\cW^{(g)}_{m+k}$, and the arguments of $\cW^{(g)}_{m+k}$ are put in bijection with the edges attached to $v$ in such a way that the distinguished one corresponds to the edge that connects $v$ to $\ell_0$.
\end{itemize}
\end{lemma}

According to the definition of the weight $	{\mathsf{w}}_{\cW'_n}(\Gamma)$, Equation~\eqref{eq:NewGraphicalFormula-cW} represents $\cW^{\np,\langle d \rangle}_n$ as a finite linear combination of $\cW^{(g)}_m(z,u;z_1',\dots,z_m')$ with $2g+m\leq d+1$ with some operators (expressed as a finite sum over graphs) in the variables $z_1',\dots,z_m'$. Note that each of these $\cW^{(g)}_m(z,u;z_1',\dots,z_m')$ satisfies the loop equations, that is,
for any  \xvital{} \nonregular{} point $q\in\cP$ such that both $dx$ and $dy$ are holomorphic at this point and $dy\ne0$ and for any $k\ge0$ we have:
\begin{equation} \label{eq:(k-1)-loop-proof}
	[u^k]e^{u y}\cW^{(g)}_n(z,u)\in\Xi_q,
\end{equation}
where $y$ is any local primitive of $dy$ (here we use the standard setup). Moreover, this relation holds identically in $z_1,\dots,z_n$. Therefore, we have \eqref{eq:(k-1)-loop-proof} in $z$ for any operator in the variables $z_1',\dots,z_m'$ applied to $\cW^{(g)}_m(z,u;z_1',\dots,z_m')$. Thus $\cW^{\np,\langle d \rangle}_n$ is expressed as a finite linear combination of elements of $\Xi_q$. Hence,
$[u^k]e^{uy}\cW^{\np,\langle d \rangle}_n\in \Xi_q$. This completes the proof of Theorem~\ref{thm:loop-equations}.

\subsection{Proof of the projection formula} In this Section we prove Theorem~\ref{thm:projectionproperty}. Consider Equation~\eqref{eq:GraphicalFormulaomega}, replacing $n$ by $n+1$. The sum over graphs in this expression can be split into two parts, depending on whether or not the leaf $\ell_{n+1}$ is attached to a $(0,1,1)$-$\omega$-vertex further attached to a $\theta$-vertex.

In the first case, since $\frac{1}{2\pi\ii}\int_\fB B(z_{n+1},\cdot)=\eta(z_{n+1})$,  we obtain the sum over graphs that is manifestly just
\begin{align}
\eta(z_{n+1})\partial_w \sum_{\Gamma\in G_{\omega_n}} \mathsf{w}_\omega(\Gamma) = \eta(z_{n+1})\partial_w \omega^\np_n.	
\end{align}
 The second sum for each $\omega^{\np,\langle d \rangle}_{n+1}$, $d\geq 0$, $n\geq 0$, is a finite linear combination of the differentials $\omega^{(g)}_{m+1}(z_1',\dots,z_m',z_{n+1})$ with $2g-1+m\geq 0$, with some operators in $z_1',\dots,z_m'$ applied to them. Since each $\omega^{(g)}_{m+1}(z_1',\dots,z_m',z_{n+1})$ has vanishing $\fA$-periods in $z_{n+1}$, we conclude that the same holds for $\omega^{\np,\langle d \rangle}_{n+1}- \eta(z_{n+1})\partial_w \omega^{\np,\langle d \rangle}_n$. This completes the proof of Theorem~\ref{thm:projectionproperty}.

\subsection{Proof of the \texorpdfstring{$x-y$}{x-y} swap relation} In this Section we prove Theorem~\ref{thm:xyswap}. Recall that next to $\omega^{(g)}_n$ and
$\omega^{\vee,(g)}_n$ we also considered in~\cite{ABDKS1} the two-index differentials $\omega^{(g)}_{m,n}$ such that $\omega^{(g)}_{m,0} = \omega^{(g)}_m$ and
$\omega^{(g)}_{0,n} = \omega^{\vee,(g)}_n$. These differentials are introduced by an inductive procedure, but it follows by their construction that they can also be defined by a closed formula
\begin{align} \label{eq:two-index}
	\omega_{m,n}=(-1)^n
	\left(\prod_{i=m+1}^{m+n}\sum_{r=0}^\infty \Bigl(-d_i\frac{1}{dy_i}\Bigr)^{r}[u_i^r]\right)
	\bW_{m,n},
\end{align}
where
\begin{align}
	\bW_{m,n}=\left(\prod_{i=1}^m\restr{u_i}{0}\right) \bW_{m+n}.
\end{align}
Now, applying the same resummation as we did above in Lemma~\ref{lem:resummation} to the Borot--Eynard formula given in Proposition~\ref{prop:Graphs-Omega}, we obtain the
following formula for $\bW^\np_n$:

\begin{lemma} We have
	\begin{equation}
		\bW^\np_n =
		\sum_{\Gamma\in G_{\bW_n}} {\mathsf{w}}_{\bW_n}(\Gamma),
	\end{equation}
	where $G_{\bW_n}$ is the set of 3-level graphs $\Gamma$ with some additional structure described as follows:
	\begin{itemize}
		\item The set $V_\ell(\Gamma)$ consists of $n$ ordered leaves $\ell_1,\dots,\ell_n$, that is, there is a fixed bijection $\set n \to V_\ell(\Gamma)$ such that $i\mapsto
		\ell_i$.
		\item There is exactly one edge attached to each $\ell_i$.
		\item 	The edge $e\in E_{\ell-\omega}(\Gamma)$ attached to $\ell_i$ is decorated by the operator $O(e)\coloneqq \restr {} {(z_i,u_i)}$.
		\item Each $(g,m,k)$-$\omega$-vertex $v$ is decorated by $\omega(v)\coloneqq \hbar^{2g-2+m+k}\bW^{(g)}_{k,m}$, and the last $m$
		(respectively, first $k$) arguments of $\bW^{(g)}_{k,m}$ are put in bijection with the edges in $E_{\ell-\omega}(\Gamma)$ (respectively, in $E_{\omega-\theta}(\Gamma)$)
		attached to $v$.
	\end{itemize}
	
\end{lemma}

Now, consider the following expression:
\begin{align} \label{eq:intermideate-x-y}
	\widetilde{\omega}^{\np,\vee}_n \coloneqq (-1)^n
	\left(\prod_{i=1}^n\sum_{r=0}^\infty \Bigl(-d_i\frac{1}{dy_i}\Bigr)^{r}[u_i^r]\right) \bW_n^\np.
\end{align}
The theory of $x-y$ duality in the standard perturbative case, in particular the formulas for the two-index differentials that we recall above in Equation~\eqref{eq:two-index}, implies that the differentials $\widetilde{\omega}^{\np,\vee}_n$ defined by this formula 
are given by
exactly the same sum over decorated graphs as $\bW^\np_n$, with the following two modifications:
\begin{itemize}
	\item Now each $(g,m,k)$-$\omega$-vertex $v$ is decorated by $\omega(v)\coloneqq \hbar^{2g-2+m+k}\omega^{(g)}_{k,m}$, and the last $m$ (respectively, the first $k$)
	arguments of $\omega^{(g)}_{k,m}$ are put in bijection with the edges in $E_{\ell-\omega}(\Gamma)$ (respectively, in $E_{\omega-\theta}(\Gamma)$) attached to $v$.
	\item The edge $e\in E_{\ell-\omega}(\Gamma)$ attached to $\ell_i$ is decorated by the operator $O(e)\coloneqq \restr {} {z_i}$.
\end{itemize}
It is not yet quite the formula for ${\omega}^{\np,\vee}_n$ that one would expect applying Corollary~\ref{cor:Graphs-omega} in the dual case. Indeed, according to Corollary~\ref{cor:Graphs-omega} the expected decoration of a $(g,m,k)$-$\omega$-vertex $v$ in a graphical formula for ${\omega}^{\np,\vee}_n$ must be $\hbar^{2g-2+m+k}\omega^{\vee,(g)}_{m+k} = \hbar^{2g-2+m+k}\omega^{(g)}_{0,m+k}$.

In order to fix this discrepancy, we notice the following. Since the sum $\omega^{(g)}_{m+1,n}(z)+\omega^{(g)}_{m,1+n}(z)$ is $d$-exact (here $z$ is the argument that changes the side in the
two-index differentials), we have
\begin{align} \label{eq:B-integral-duality}
\oint_{z\in\fB_i} \omega^{(g)}_{m+1,n} = 	-\oint_{z\in\fB_i} \omega^{(g)}_{m,1+n}, \qquad i=1,\dots,\cg.
\end{align}
Thus 
\begin{align}
\oint_{z_1\in\fB_{i_1}} \cdots \oint_{z_k\in\fB_{i_k}} \omega^{(g)}_{k,m} & = 	(-1)^k\oint_{z_1\in\fB_{i_1}} \cdots \oint_{z_k\in\fB_{i_k}} \omega^{(g)}_{0,m+k}
\\ \notag 
&=(-1)^k\oint_{z_1\in\fB_{i_1}} \cdots \oint_{z_k\in\fB_{i_k}} \omega^{(g),\vee}_{m+k}, \qquad 1 \leq i_1,\dots,i_k \leq \cg,
\end{align}
where the integrals are taken with respect to the first $k$ arguments. This implies that in the description of $\widetilde{\omega}^{\np,\vee}_n$ we can replace the decoration of a $(g,m,k)$-$\omega$-vertex $v$ by $\hbar^{2g-2+m+k}\omega^{\vee,(g)}_{m+k}$ instead of $\hbar^{2g-2+m+k}\omega^{(g)}_{k,m}$, with an overall factor of $(-1)^K$, where $K=\sum k(v)$ is the total number of edges in $E_{\omega-\theta}(\Gamma)$. This factor can be absorbed by replacing the operator $\partial_w$ with $-\partial_w$ on each of these edges. Applying $(-\partial_w)^K$ to the vertex decoration $\log\Theta_*(w)$ is equivalent to applying $\partial_u^K$ to $\log\Theta_*(-u)$ evaluated at $u=-w$. Since $\Theta_*$ is an odd function, $\log\Theta_*(-u) = \log\Theta_*(u) + \mathrm{const}$. Since each $\theta$-vertex is acted upon by at least one derivative, this constant difference vanishes. This identifies $\widetilde{\omega}^{\np,\vee}_n$ defined by Equation~\eqref{eq:intermideate-x-y} with the formula for ${\omega}^{\np,\vee}_n$ given by Corollary~\ref{cor:Graphs-omega} applied in the dual case and evaluated at $-w$.

This proves the first assertion of Theorem~\ref{thm:xyswap}. Now, according to~\cite{ABDKS1}, the second statement is a formal inverse of the first one (or,
alternatively, one can prove it by exactly the same argument reversing the roles of $dx$ and $dy$).

\subsection{Proof of the deformation formula} In this section we prove Theorem~\ref{thm:Deformation-y}. We just substitute Equation~\eqref{eq:GraphicalFormulaomega} on the left hand side and on the right hand side of Equation~\eqref{eq:Deformation-y}.

On the left hand side, we can combine the known deformation formula for $\omega^{(g)}_n$ (see e.g. \cite{alexandrov2024topologicalrecursionrationalspectral}) and the Leibniz rule. This means that we change the decoration of one of the $(g,m,k)$-$\omega$-vertices with $2g-2+m+k>0$ from $\omega^{(g)}_{m+k}$ to $\sum_{q\in\cP} \res_{z=q} \omega^{(g)}_{m+k+1}(z) \int_q^{z} \Delta y\,dx$.

On the right hand side, there are three possible cases:
\begin{enumerate}
	\item First possible situation is that the leaf corresponding to the extra variable $z$ is connected to a $(g,m,k)$-$\omega$-vertex with $2g-2+m+k>1$. The sum over all possible cases like this coincides with the left hand side.
	\item Second possibility is that the leaf corresponding to the extra variable $z$ is connected to a $(0,m,k)$-$\omega$-vertex with $m+k=3$. But then notice that $ \int_q^{z} \Delta y\,dx$ has a double zero at $q$ and $\omega^{(0)}_3$ at most a double pole at $q$. This means that the sum of residues  $\sum_{q\in\cP} \res_{z=q} \omega^{(0)}_{3}(z) \int_q^{z} \Delta y\,dx$ vanishes and these terms do not contribute.
	\item The third possibility is that the leaf corresponding to the extra variable is connected to a $(0,1,1)$-$\omega$- or $(0,2,0)$-$\omega$-vertex . But again, $\omega^{(0)}_2=B$ is holomorphic at each $q\in \cP$, hence the sum of residues  $\sum_{q\in\cP} \res_{z=q} \omega^{(0)}_{2}(z) \int_q^{z} \Delta y\,dx$ vanishes and these terms do not contribute.
\end{enumerate}

Thus we prove that the left hand side and the right hand side of Equation~\eqref{eq:Deformation-y} are equal. This completes the proof of the theorem.

\subsection{Proof of the KP integrability}\label{sec:proofKP}

In this Section we prove Theorem~\ref{thm:KP-integrable}.

\subsubsection{Standard setup} \label{sec:standardsetup}
We assume first that we are in the realm of the standard setup of topological recursion. That is, we assume that $\cP$ is the set of zeros of $dx$ and $dy$ is regular and non-vanishing at each point $q\in \cP$.

Note that both the $x-y$ swap and deformations of $dy$, such that $dy$ is regular at zeros of $dx$, preserve the KP integrability of a system of differentials, see~\cite[Theorem 2.8]{ABDKS3} and~\cite[Corollary 2.6]{alexandrov2024topologicalrecursionrationalspectral}, respectively. An important extra observation is that these statements can be directly applied to $\{\omega^{\np}_{n}\}$. Indeed, the $x-y$ swap relation for the differentials $\{\omega^{\np}_{n}\}$ is described by a combination of \eqref{eq:incl-excl-np},\eqref{Wom},\eqref{eq:Omeganp-intermsof-omeganp}, and \eqref{eq:xyrel},
which, up to a slightly different convention (in particular, note that $\Omega$'s and $\bW$'s in \cite{ABDKS3} are not differentials, but corresponding functions) and a substitution $w \mapsto -w$, coincides with the $x-y$ swap relation for the differentials $\{\omega_{n}\}$ of \cite[Section 4.1]{ABDKS3}. Then the proof of \cite[Theorem 2.8]{ABDKS3} can be applied to a collection of differentials $\{\omega^{\np}_{n}\}$ without any changes. We conclude that the original system of differentials $\{\omega^{\np}_{n}\}$ evaluated at $w$ is KP integrable if and only if the dual system of differentials $\{\omega^{\np,\vee}_{n}\}$ evaluated at $-w$ is KP integrable. Since KP integrability holds for generic values of the parameter, integrability at $-w$ is equivalent to integrability at $w$.
The formula of Theorem~\ref{thm:Deformation-y} is equivalent to \cite[Equation (22)]{alexandrov2024topologicalrecursionrationalspectral}. According 
to \cite[Corollary 2.6]{alexandrov2024topologicalrecursionrationalspectral} the latter equation describes a deformation that preserves the KP integrability of a collection of differentials. Therefore, the deformation given by Equation~\eqref{eq:Deformation-y} preserves the KP integrability of $\{\omega^{\np}_{n}\}$.

 Combining these two operations, we are also allowed to deform $dx$ in such a way that $dx$ remains regular at the zeros of $dy$.
Thus, our strategy is to apply a sequence of deformations of $dx$ and $dy$ such that at the end we arrive to the situation that $dx$ has poles of sufficiently high order at all zeros of $dy$. This would mean that the system of dual differentials $\{\omega^{\vee,(g)}_n\}$ is computed by generalized topological recursion with $\cP^\vee = \emptyset$, and all differentials in the stable range are equal to zero: $\omega^{\vee,(g)}_n = 0$ for $2g-2+n>0$. Hence, by Lemma~\ref{lem:omega-np-analytic}, $\omega^{\np,\vee}_n = \omega^{\Kr}_n$, and this system is integrable. Hence, its $x-y$ dual system of differentials $\{\omega^{\np}_n\}$ is also KP integrable, as well as any other system of non-perturbative differentials connected to it via KP integrable deformations, including the original system of non-perturbative differentials.

Let us describe a construction of the desired deformation. Denote the initial differentials by $dx_0$ and $dy_0$ and assume that $dx_0$ is regular and non-vanishing at the zeros of $dy_0$. We need to choose some $dx_1$ and $dy_1$ satisfying the following conditions:
\begin{itemize}
	\item $dy_0$ is regular and non-vanishing at the zeros of $dx_1$;
	\item $dy_1$ is regular and non-vanishing at the zeros of $dx_1$;
	\item $dx_1$ is regular and non-vanishing at the zeros of $dy_0$;
	\item $dx_1$ has poles of sufficiently high order at the zeros of $dy_1$.
\end{itemize}
The choice is done in the following order. First, we choose some $dy_1$ whose set of zeros is disjoint from the set of zeros of $dy_0$. Then we choose $dx_1$ such that it has poles of sufficiently high order at the zeros of $dy_1$ and no other poles, and its zeros are disjoint from the zeros and poles of $dy_0$ and the poles of $dy_1$. Both choices are obviously possible.

Now we first deform $(dx_0,dy_0)$ to $(dx_t,dy_0)$, $0\leq t \leq 1$, where $dx_1$ is the one we have chosen, and in order to avoid any problems  with the existence of the global differentials $dx_t$ for $0<t<1$ we demand only that they are defined locally at the zeros of $dy_0$ (this relaxed setup is sufficient for the integrability properties, cf.~\cite[Section 4]{alexandrov2024topologicalrecursionrationalspectral}). Then we deform $(dx_1,dy_0)$ to $(dx_1,dy_s)$, $0\leq s \leq 1$, where $dy_1$ is the one we have chosen, and $dy_s$ for $0<s<1$ are defined only locally at the zeros of $dx_1$.

As we discussed above, this sequence of deformations preserves KP integrability, and the resulting system of differentials satisfies the conditions  for KP integrability discussed above. This completes the proof of Theorem~\ref{thm:KP-integrable} in the standard setup of topological recursion.

\subsubsection{Extension to generalized setup} Now assume that the input data $(\Sigma, dx, dy, \cP)$ corresponds to generalized topological recursion in the sense of~\cite{alexandrov2024degenerateirregulartopologicalrecursion}. Then we can use exactly the same argument as in~\cite[Proof of Theorem 6.4]{alexandrov2024degenerateirregulartopologicalrecursion}, which, in a nutshell, says that the system of differentials of generalized topological recursion can be realized as the limit of a one-parameter family of a system of the differentials of the standard topological recursion. This extends without any effort to the non-perturbative systems of differentials. The observation that the KP integrability property is a closed condition completes the proof in this case.

\section{Back to the Borot-Eynard formulation of KP integrability} 
\label{sec:discussion-formulations}

The goal of this Section is to briefly recall how the concept of KP integrability is treated in~\cite{BorEyn-AllOrderConjecture} and compare our results to the conjectures in~\emph{op.~cit.}. First of all, notice that our Theorem~\ref{thm:KP-integrable} coincides with~\cite[Theorem 8.1]{BorEyn-AllOrderConjecture}, which is a conditional statement in~\emph{op.~cit.} subject to~\cite[Conjecture 7.1]{BorEyn-AllOrderConjecture}. Indeed, the kernel derived in \cite[Propoposition~8.1]{BorEyn-AllOrderConjecture} coincides at each fixed spectral curve data with  $\Omega^{\np}_1$. Thus,  Equation~\eqref{eq:KP-integrable-smallomega} reproduces the statement of~\cite[Theorem 8.1]{BorEyn-AllOrderConjecture}. 

However, note that Theorem~\ref{thm:KP-integrable} is proved for any input data of topological recursion, and thus can be considered in families. This is sufficient to revert the argument of~\cite{BorEyn-AllOrderConjecture} and to directly derive the initial form of their conjecture. Let us explain this. Throughout this section we assume the generalized setup for topological recursion.

Let $o$ be a regular point for the system of non-perturbative differentials and $z$ a local coordinate at $o$. Let $\tau_{o,z}^{\np}$ be a formal power series in $t_1,t_2,\dots$ with the constant term equal to $1$ defined by \eqref{eq:n-differential} for $\omega_n^{\np}$. The formal variables $t_1,t_2,\dots$ provide what might be called the Krichever parameterization of the formal neighborhood of the initial data of topological recursion (cf.~\cite[Section 4]{BorEyn-AllOrderConjecture}), but this interpretation is not important for our analysis. 

Using~\cite[Section 3.2]{ABDKS3} we immediately have the following corollary:

\begin{corollary} For any regular point $o$ and a local coordinate $z$ at  $o$ the tau function $\tau_{o,z}^{\np}$ is a tau function of the KP hierarchy. 
\end{corollary}

Let us also give a formula for $\log \tau_{o,z}^{\np}$ in terms of the 3-level graphs. Let $G_{\log\tau}$ be the set of 3-level graphs $\Gamma$ with some additional structure described as follows:
\begin{itemize}
	\item The set $V_\ell(\Gamma)$ consists of $n\geq 1$ unordered leaves $\ell_1,\dots,\ell_n$ (that is, the leaves can be permuted by the automorphisms of the graph). Here $n$ is not fixed, it is an arbitrary positive integer.
	\item There is exactly one edge attached to each $\ell_i$.
	\item Each edge $e\in E_{\ell-\omega}(\Gamma)$ is decorated by the operator $O(e)\coloneqq \res_{z=0} \sum_{k=1}^\infty \frac{t_k}{z^{k}}$.
	\item For $(g,m,k)\not=(0,2,0)$, each $(g,m,k)$-$\omega$-vertex $v$ is decorated by $\omega(v)\coloneqq \hbar^{2g-2+m+k}\omega^{(g)}_{m+k}$, and the arguments of $\omega^{(g)}_{m+k}$ are put in bijection with the edges attached to $v$.
	\item The $(0,2,0)$ vertex is decorated by $B(z_1,z_2)-\frac{dz_1dz_2}{(z_1-z_2)^2}$. 
\end{itemize}

\begin{proposition} \label{prop:tau_graph}
We have
\begin{align}
	\log \tau_{o,z}^{\np} = %
	\sum_{\Gamma\in G_{\log\tau}} \mathsf{w}_{\log\tau}(\Gamma).
\end{align}
Alternatively, one can rewrite it as (cf.~\cite[Eq.~(99)]{BorEyn-knots})
\begin{align}\label{eq:Krtau-np}
	\tau_{o,z}^{\np} & = \dfrac{\exp \left({\frac12\sum_{k,l\ge1}b_{k,l}t_kt_l}\right)}{\exp\left(\sum_{\scriptstyle m\geq 1}\frac{1}{m!} \Big(\prod_{j=1}^m \frac{1}{2\pi\ii}\oint_{z_j\in \fB} {\partial_w}\Big)  \omega_m (z_{\set{m}})\right) \,\Theta_*(w) } \\ \notag 
	& \phantom{=} \times
	\exp \left(\sum_{\substack{m,k\geq 0\\ m+k>0}}\frac{1}{m!k!} \left(\prod_{i=1}^m\res_{z_i=0} \sum_{l=1}^\infty \frac{t_l}{z_i^{l}}\right) \right.
	\\ \notag & \left. \qquad \qquad \qquad 
	 \left(\prod_{j=m+1}^{m+k}\frac{1}{2\pi\ii}\int_{z_j\in\fB} \partial_w\right) \omega_{m+k}(z_{\set{m+k}})\right)
	\Theta_*\Big(\sum_{k=1}^\infty t_k U_k+w\Big),
\end{align}
where the constants $U_k$ and $b_{k,l}$ are defined in~\eqref{eq:Uka} and
\eqref{eq:Bz12}.
\end{proposition}

The proof is straightforward by analogy with Proposition~\ref{prop:Graphs-Omega} above. Note that if the topological recursion is trivial, then by Remark~\ref{rem:Graphs-02}, in particular by Equation~\eqref{eq:theta-shift-Krichever}, we obtain the original Krichever formula as in Equation~\eqref{eq:Krtau}. 

Now we can go through our construction keeping track of the variables $t_1,t_2,\dots$. In particular, we introduce
\begin{align} \label{eq:omega-np-t}
	\omega^{\mathrm{np}}_n(z_{\set{n}} \, |\, t_1,t_2,\dots) \coloneqq \delta_{z_1}\cdots \delta_{z_n} \log \tau_{o,z}^{\np} +  \delta_{n,2}\frac{dz_1dz_2}{(z_1-z_2)^2}
\end{align}
in the neighborhood of the point $o$ and then extend them analytically to be formal power series in $t_1,t_2,\dots$ with the coefficients being globally defined meromorphic differentials on $\Sigma^n$ (note that now we don't set $t\to0$ as in \eqref{eq:n-differential}). Note that it is not obvious that $\omega^{\mathrm{np}}_n(z_{\set{n}} \, |\, t_1,t_2,\dots)$ defined by~\eqref{eq:omega-np-t} can indeed be analytically extended. To this end, one can use an alternative equivalent formula 
\begin{align} \label{eq:omega-np-t-eq}
	& \omega^{\mathrm{np}}_n(z_{\set{n}} \, |\, t_1,t_2,\dots) = 
	\\ \notag &
	\sum_{m=0}^\infty \frac{1}{m!} \left(\prod_{i=n+1}^{n+m} \res_{z_i=0}\sum_{k=1}^\infty \frac{t_k}{z_i^{k}}\right) \bigg(\omega^{\mathrm{np}}_{n+m}(z_{\set{n+m}}) - \delta_{n,1}\delta_{m,1} \frac{dz_1dz_2}{(z_1-z_2)^2} \bigg),
\end{align} 
which gives a globally defined $n$-differential.

Extend all further definitions to retain this dependence on $t_1,t_2,\dots$, in particular, defined $\Omega^{\np,\bullet}_n (z^+_{\set n},z^-_{\set n}  \, |\, t_1,t_2,\dots)$. A direct corollary of Theorem~\ref{thm:KP-integrable} is the following

\begin{corollary}\label{cor:KP-integrable} The system of differentials $\{\omega^{\np}_n(z_{\set{n}} \, |\, t_1,t_2,\dots)\}$ is KP integrable, that is, we have:
	\begin{align} \label{eq:KP-integrable-smallomega-t1t2}
		\omega^{\np}_n (z_{\set n} \, |\, t_1,t_2,\dots) & = {\det}^\circ (\Omega^\np_1 (z_i,z_{j}  \, |\, t_1,t_2,\dots)),\quad n\ge2.
		\\	\label{eq:KP-integrable-Bigomega-t1t2}
		\Omega^{\np,\bullet}_n (z^+_{\set n},z^-_{\set n} \, |\, t_1,t_2,\dots) & = {\det} (\Omega^\np_1 (z^+_i,z^-_{j} \, |\, t_1,t_2,\dots)).
	\end{align}
\end{corollary}

Note that once we decided to retain the dependence on $t_1,t_2,\dots$, we get an additional equation for $\omega^\np_n$, namely, 
\begin{align}
	\omega^\np_{n+1}(z_{\set{n+1}} \, |\, t_1,t_2,\dots) = \delta_{z_{n+1}}  \omega^\np_{n}(z_{\set{n}} \, |\, t_1,t_2,\dots) +  \delta_{n,1}\frac{dz_1dz_2}{(z_1-z_2)^2}.
\end{align}
Hence, 
\begin{align}
	\delta_z \Omega_1^{\np}(z_1^+,z_1^-\, |\, t_1,t_2,\dots) = \restr{z_2^+}{z}\restr{z_2^-}{z}\Omega_2^{\np}(z_{\set{2}}^+,z_{\set{2}}^-\, |\, t_1,t_2,\dots).
\end{align}
On the other hand, Equation~\eqref{eq:KP-integrable-Bigomega-t1t2} implies that 
\begin{align}
	\Omega_2^{\np}(z_{\set{2}}^+,z_{\set{2}}^-\, |\, t_1,t_2,\dots) = - \Omega_1^{\np}(z_1^+,z_2^-\, |\, t_1,t_2,\dots) \Omega_1^{\np}(z_2^+,z_1^-\, |\, t_1,t_2,\dots).
\end{align}
Thus, we obtain the following direct corollary:
\begin{corollary} We have 
	\begin{align}
		\delta_z \Omega_1^{\np}(z_1^+,z_1^-\, |\, t_1,t_2,\dots) = 
		- \Omega_1^{\np}(z_1^+,z\, |\, t_1,t_2,\dots) \Omega_1^{\np}(z,z_1^-\, |\, t_1,t_2,\dots).
	\end{align}
\end{corollary}
This reproduces~\cite[Conjecture 7.1]{BorEyn-AllOrderConjecture}, written in formal Krichever's parameterization of the spectral curve data in the neighborhood of a given spectral curve. In particular, in the case of a trivial topological recursion, the argument above derives a version of the trisecant Fay identities from the KP integrability of Krichever differentials. 

\printbibliography

\end{document}